# Controlling Flow Separation over a Curved Ramp Using Vortex Generator Microjets

Mohammad Javad Pour Razzaghi[1], Yasin Masoumi[2], Seyed Mojtaba Rezaei Sani[3], Guoping Huang[1]*

[1]*College of Energy and Power Engineering, Nanjing University of Aeronautics and Astronautics, Nanjing, 210016, PR China*

[2]*Acoustics Research Laboratory, Center of Excellence in Experimental Solid Mechanics and Dynamics, School of Mechanical Engineering, Iran University of Science and Technology, Narmak, Tehran 16846-13114, Iran*

[3]*Department of Physics, North Tehran Branch, Islamic Azad University, Tehran 16511-53311, Iran*

*hgp@nuaa.edu.cn

**Abstract:**

Introducing a fluid microjet into the boundary layer to increase fluid momentum and hence delay separation is a method for actively controlling a flow separation region. The present work numerically analyzed the control of a separation bubble behind a ramp. For this purpose, we first verified the numerical results for a flow (without a jet) over the ramp against reliable experimental studies from the literature. Next, the effects of introducing a microjet to the flow were also verified. A jet was then placed at three different distances above the ramp to study its impact on various parameters, including velocities, Reynolds stresses, pressure, vorticity, streamlines, and the separation bubble size. As the jet was moved further back, the jet-induced upwash region grew considerably. Finally, the effects of using three identical jets were studied and compared against those of a single jet. The results indicated that using a three-jet array shrank the separation bubble. Using an array with $d/D$ = 15 can limit laterally the separation bubble about 2.75 times smaller than a single jet in the $z$-direction. Also, the employment of the jet managed to decrease the length of the separation zone in the $x$-direction up to 78%, in the case of $L_x/L_1$ = 0.0143 and $d/D$ = 10.

## 1. Introduction

An adverse pressure gradient occurs when the static pressure increases in the direction of the flow. So, the boundary layer size increases directly above the wall while the boundary layer velocity decreases [1]. For a large enough adverse pressure gradient, the direction of the boundary layer velocity changes, separating the boundary layer [2,3]. This local phenomenon is accompanied by the stall (the creation of large vortices, an increase in drag, and a lift decrease). Some flow control methods can delay the location of this separation [3].

Flow control methods are divided into passive (no external energy source is required) and active (an external energy source is needed). The energy is transferred from the main flow to the boundary layer in the passive flow control method. Also, the energy is transferred from some external energy source to the fluid in the active flow control method [1-3]. One of these active control methods is the vortex generator jet (VGJ) method. This technique resembles a well-known method of using a small vane vortex generator [4]. VGJs have apparent advantages over vane vortex generators; for example, they do not suffer from drag penalties and can be used in the active flow control system. However, experimental or computational fluid dynamics (CFD) modeling of the flow with VGJs is more complex and costly than vane vortex generators [5].

Further, instead of a single jet, an array of jets can be used, boosting the vortex forming in the flow and extending the effects further into the flow. In a single jet configuration, the turbulent boundary layer evolves according to the skew angle of the jet's direction as longitudinal vortices merge or move independently. Therefore, this property of microjets in active flow control can be utilized in various applications, including boundary layer separation. For this reason, they have been extensively studied from different aspects. In the following, a comprehensive review of this issue was presented.

McCurdy [6] was possibly the first researcher to investigate the effects of vortex generators (VGs) on boundary layer flow control on an airfoil in 1948. A few years later, in 1952, Wallis [7] compared an inclined active VG jet (VGJ) with an inclined passive vane VG (VVG). According to his investigation, VGJ arrays had a similar impact on a flow as an array of VVGs. Pearcey [4] was the first researcher to investigate the results of using the VVGs and the VGJs for boundary layer separation control. Simpson et al. [8] investigated turbulent boundary layer separation problems due to an adverse pressure gradient. These studies were primarily aimed at the experimental analysis of the average velocity and Reynolds stresses in the downwash and upwash regions. In1996, Webster et al. [9] investigated turbulence characteristics of a boundary layer over a two-dimensional bump by an experimental method. They analyzed velocity, Reynolds stress, and pressure along the flow. Song and Eaton [10] experimentally studied Reynolds number effects on a turbulent boundary layer with separation, reattachment, and recovery. They investigated velocity and Reynolds stress at various cross-sections of the flow.

In 2006, Radhakrishnan et al. [11] considered various numerical turbulent models to study the fluid flow over an asymmetric domain. They analyzed the flow over a ramp and a bump, considered Spalart-Allmaras Reynolds-averaged Navier-Stokes (SA-RANS), $k$-$\varepsilon$, shear stress transport (SST), and wall-modeled large eddy simulation (WMLES) methods to predict separation and reattachment points, and compared the results from these methods to experimental results. Von Stillfried et al. [12] reported that the actual boundary layer conditions and many combinations are possibly essential in the optimizing parameters of VGJs. According to this investigation, optimum ranges for pitch and skew angles are between 15-45° and 90-135°, respectively. Bentaleb et al. [13] investigated large eddy simulations of turbulent boundary layer separation from a rounded step. Apart from offering insight into the physics of separation, their study constitutes a valuable data set for benchmarking model solutions and investigating statistical turbulence closure proposals. In 2014, Lasagna et al. [14] investigated stream-wise vortices originating from the interaction of a synthetic jet and a turbulent boundary layer by an experimental method. According to their results, the leeward vortex intensified while the other became weaker by increasing the slot yaw angle. Moreover, the vortices grew in size and intensity by increasing the jet velocity ratio (VR) and the slot yaw angle.

Feng et al. [15] studied a model of the trajectory of an inclined jet in incompressible crossflow. Also, they investigated the effect of jet entrainment. The horseshoe and wake vortices may create a low-pressure region on the wall. This may alter the jet trajectory and influence the vortex circulation in low-velocity ratios and skew angle near 90°. Szwaba et al. [16] investigated the influence of rod vortex generators on a flow pattern downstream using experimental and numerical methods in 2019. They put out a rod instead of a jet to show that the application of a rod can introduce the same effect as a jet and so introduced a new flow control method dedicated mainly to external flows. In another study, Liu et al. [17] used the compression corner calculation model and conducted detailed numerical investigations in the supersonic flow field. They studied the effects of different injection pressure ratios, various actuation positions, and different nozzle types. Furthermore, they mentioned that the distance between the counter-rotating vortex pair and the wall surface is an essential factor. Wang and Ghaemi [18] studied the effects of vane sweep angle on the vortex generator wake. Their experimental studies compared the stream flowing over a pair of deltas, trapezoidal, and rectangular vanes. They also discussed such parameters as vortex spacing, intensity, and turbulence.

However, there are still fewer studies on minute vortex generator jets (micro-VGJs) and the interaction of vortices induced by microjets. Suryadi [19] proposed a similarity model for separated turbulent boundary layers as a function of the outer flow parameters, the velocity associated with the maximum local mean shear, and the reduced wall-normal coordinate. Shu and Li [20] investigated the separated and reattaching flows over a blunt flat plate with different leading-edge shapes using particle image velocimetry and surface pressure measurements. This study altered some geometric parameters and analyzed how flow velocity, pressure, and vorticity were affected. Recently, John et al. [21] numerically analyzed the $k$-$\varepsilon$, $k$-$\omega$, Reynolds stress model (RSM), SST $k$-$\omega$, and LES models of a wind tunnel and studied the flow velocity and turbulence parameters in different cross-sections of the tunnel. In another 2020 study, Zhang et al. [22] analyzed the $k$-$R$ turbulence model for wall-bounded flows. They compared the results from this model and those from the direct numerical simulation (DNS), $k$-$\varepsilon$, and SST $k$-$\omega$ models and found that they all predicted almost similar results. However, the results from their proposed model followed the experimental results more closely under specific conditions. In the same year, Gorbushin et al. [23] measured the mean parameters of a two-dimensional incompressible turbulent boundary layer with zero pressure gradient on the smooth wall of a wind tunnel. In their paper, the main objective was to examine and compare the results of their model with others.

Very recently, Lu et al. [24] numerically simulated spatially developing incompressible turbulent boundary layers (SDTBLs) with adverse pressure gradient (APG) over two-dimensional smooth curved ramps by solving the RANS equations. They analyzed pressure along the flow by changing various parameters. In our previous study [25], the impact of microjets on a turbulent boundary layer was numerically investigated. We studied the effects of changing the microjet angle and velocity ratio on flow separation and proposed 30° and 60° for pitch and skew angles as the most effective angles, respectively. In another study [26], we numerically analyzed how a pair of microjets affected turbulent boundary layers. The distance between the jets and their angles with respect to each other were changed to analyze merging and separation flow. Moreover, we have numerically analyzed how the various turbulence models can predict the effects of a microjet on turbulent boundary layers. According to our results, it was concluded that the four-equation transitional SST model provides the best prediction [27].

Considering the above literature review, a numerical analysis of flow separation control behind a ramp is presented in what follows. For this purpose, the numerical results were verified against experimental ones from the literature for the case without a jet. The effects of introducing a jet to the flow were then studied by comparing this case's numerical and experimental results. Once the numerical results were verified, we investigated the effects of a fluid jet on flow separation behind a ramp. These effects were studied with a jet at various longitudinal distances above the ramp. The separation bubble, velocities, Reynolds stresses, pressure gradients, vorticity, and streamlines were examined for these cases. Finally, we analyzed the separation bubble region using an array of three identical jets at equal distances.

## 2. Numerical Modeling

### 2.1. Modeling Details

Fig. 1 shows the computational domain similar to that used by [10]. The figure provides the jet dimensions, position, and orientation. The computational domain was 2789 mm long, 500 mm wide, and 152 mm high. The ramp was 70 mm long and 21 mm high. The ramp geometry was a circular sector of radius 127 mm at the exit of the domain, while it was a 5th degree polynomial at the entry. The jets were placed above the ramp at various longitudinal distances ($L_x/L_l$: Jet 1 = 0.0214, Jet 2 = 0.286, and Jet 3 = 1.357). The jet was 1 mm in diameter with $\alpha = 30°$ and $\beta = 45°$. Inlet velocity and outflow boundary conditions were considered at the domain inlet and outlet, respectively. A wall boundary condition was used at the top and bottom, and a symmetric condition was applied for the two sides.

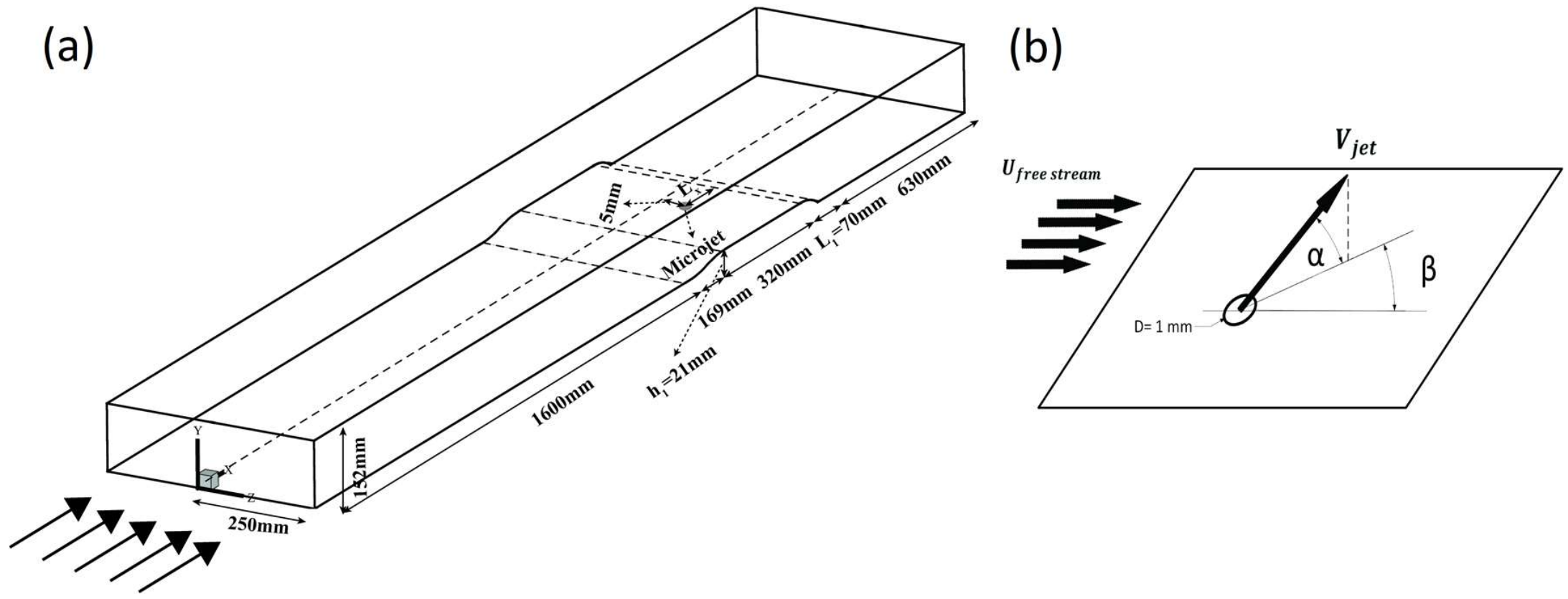

Figure 1: (a) Geometry and dimensions of the computational domain and the fluid jet. (b) Orientation of the microjet relative to the coordinate axes.

The fluid considered was the air with viscosity $\mu = 1.81 \times 10^{-5}\ kg/s.m$ and density $\rho = 1.12\ kg/m^3$. The inflow velocity was 15 m/s, and its turbulence intensity was 2%. The turbulent to molecular viscosity ratios at the inlet and outlet boundaries were set to 5% and 10%, respectively.

### *2.2. Governing Equations*

Navier-stokes equations were expanded to RANS and solved numerically for flow analysis. The finite-volume method was employed for the numerical analysis of these equations in the computational domain. Further, pressure and momentum equations were expanded by the second-order upwind scheme, coupling velocity and pressure fields by the SIMPLE algorithm. The problem involved the following continuity, momentum, and turbulence equations [28, 29]:

$$\frac{\partial \rho}{\partial t} + \frac{\partial(\rho u_i)}{\partial x_i} = 0, \tag{1}$$

$$\frac{\partial \mathrm{u}_i}{\partial \mathrm{t}} + \frac{\partial(u_i u_j)}{\partial x_j} = g_i - \frac{1}{\rho}\frac{\partial P}{\partial x_i} + \frac{\partial}{\partial x_j}\left[(\nu + \nu_t)\left(\frac{\partial u_i}{\partial x_j} + \frac{\partial u_j}{\partial x_i}\right)\right], \tag{2}$$

where $g$, $u$, and $P$ are the Earth's gravitational acceleration, velocity, and pressure, respectively. Also, $\rho$, $\nu$, and $\nu_t$ are density, dynamic viscosity, and turbulence viscosity, respectively. $t$ represents time, and position is denoted by $x$. In fact, the momentum equation establishes a relationship between inertia, pressure, volumetric, and viscosity forces. It should be noted that the turbulence dynamic viscosity, $\nu_t$, was obtained through the transition SST model. The transition SST model is based on coupling the SST *k-ω* transport equations with two other transport equations, one for the intermittency and one for the transition onset criteria, in terms of momentum-thickness Reynolds number. A proprietary empirical correlation (Langtry and Menter [30]) has been developed to cover standard bypass transition as well as flows in low free-stream turbulence environments. The transport equation for the intermittency, $\gamma$, reads:

$$\frac{\partial(\gamma)}{\partial t} + \frac{\partial(u_j \gamma)}{\partial x_j} = P_{\gamma 1} - E_{\gamma 1} + P_{\gamma 2} - E_{\gamma 2} + \frac{\partial}{\partial x}\left[\left(\nu + \frac{\nu_t}{\sigma_\gamma}\right)\frac{\partial \gamma}{\partial x_j}\right], \tag{3}$$

where the transition sources are defined as follows:

$$E_{\gamma 1} = P_{\gamma 1}\gamma \quad \text{and} \quad P_{\gamma 1} = 2F_{length} \cdot \rho S[\gamma F_{onset}]^{c_{\gamma 3}}, \tag{4}$$

where $S$ is the strain rate magnitude, and $F_{length}$ is an empirical correlation that controls the length of the transition region. The destruction/relaminarization sources are defined as follows:

$$E_{\gamma 2} = P_{\gamma 2}\gamma \quad \text{and} \quad P_{\gamma 2} = (2c_{\gamma 1})\rho\Omega\gamma F_{length}, \tag{5}$$

where Ω is the vorticity magnitude. The transport equation for the transition momentum thickness Reynolds number $\widetilde{Re_{\theta_t}}$ is:

$$\frac{\partial(\widetilde{Re_{\theta_t}})}{\partial t} + \frac{\partial(u_j\widetilde{Re_{\theta_t}})}{\partial x_j} = P_{\theta t} + \frac{\partial}{\partial x_j}\left[\sigma_{\theta t}(\nu + \nu_t)\frac{\partial \widetilde{Re_{\theta_t}}}{\partial x_j}\right]. \tag{6}$$

The source term is defined as follows:

$$P_{\theta t} = c_{\theta t}\frac{\rho}{t}\left(Re_{\theta_t} - \widetilde{Re_{\theta_t}}\right)(1 - F_{\theta t}), \tag{7}$$

where $t$ is a time scale, $\widetilde{Re_{\theta_t}}$ and $Re_{\theta_t}$are transported scaler and local value, and $F_{\theta t}$ is the blending function used to turn off the source term in the boundary layer and its value is 1.0 in the boundary layer and zero in the freestream. Also, $\sigma_{\theta t} = 2$ and $c_{\theta t} = 0.03$ are the model constants for the $\widetilde{Re_{\theta_t}}$ equation.

The transition model interacts with the SST turbulence model as follows:

$$\frac{\partial(k)}{\partial t} + \frac{\partial(u_j k)}{\partial x_j} = \widetilde{P_k} - \widetilde{D_k} + \frac{\partial}{\partial x_j}\left[(\nu + \sigma_k\nu_t)\frac{\partial k}{\partial x_j}\right], \tag{8}$$

where $\widetilde{P_k}$ and $\widetilde{D_k}$ are the new production and destruction terms for the SST model, respectively. The rationale behind the above model formulation is detailed in Ref. 30.

### *2.3. Verification and Validation*

In the following, the results for the case without a jet were first verified against the experimental results from Ref. 10. Then, the results for the case with a jet were validated by comparison against the experiment. Next, the grid-independence test was carried out with the jet included in the computational domain. Fig. 2 compares the present work's numerical results with the experimental results. The velocity and Reynolds stress were examined. The figure shows that the numerical results with less than 1% error for the velocity and about 10% for the Reynolds stress have a reasonable agreement with experimental results. Given the uncertainties pointed out in Ref. 10, these values are acceptable as errors. A comparison between the separation bubble regions is also shown in the figure, confirming the numerical scheme predicted the region reasonably well. There is less than 1% error in this case.

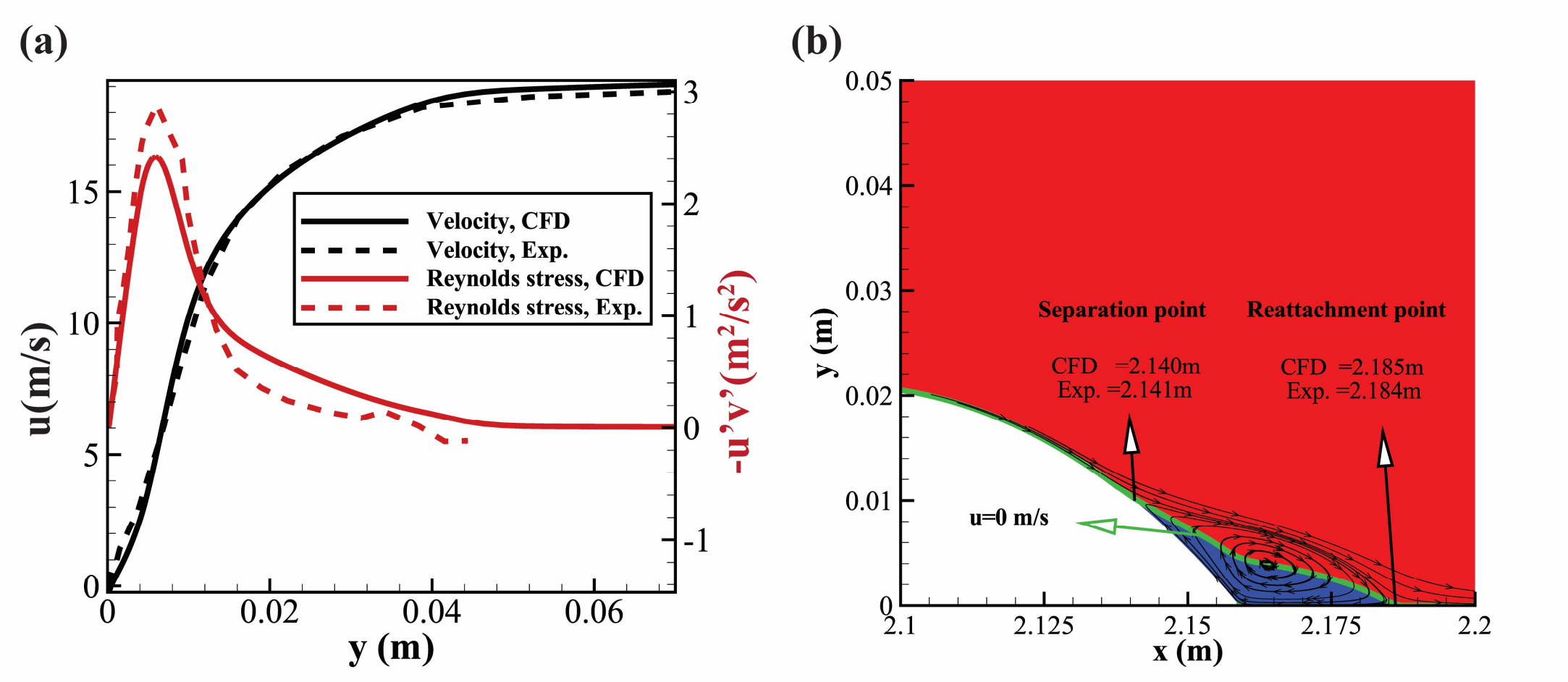

Figure 2: Comparison of numerical and experimental results [10] for the computational domain without a jet. (a) For the velocity and Reynolds stress, (b) For the separation bubble region.

Fig. 3 compares the experimental and numerical results for a surface flow with a microjet. The comparison was made at 9 cm from the microjet by considering $u/U$ for VRs = 1, 2, and 4, corresponding to percentage errors of 1%, 3%, and 5%, respectively. As the figure shows, there is good agreement between the experimental and numerical results. It was concluded that the numerical results could be used in further stages through the analysis process.

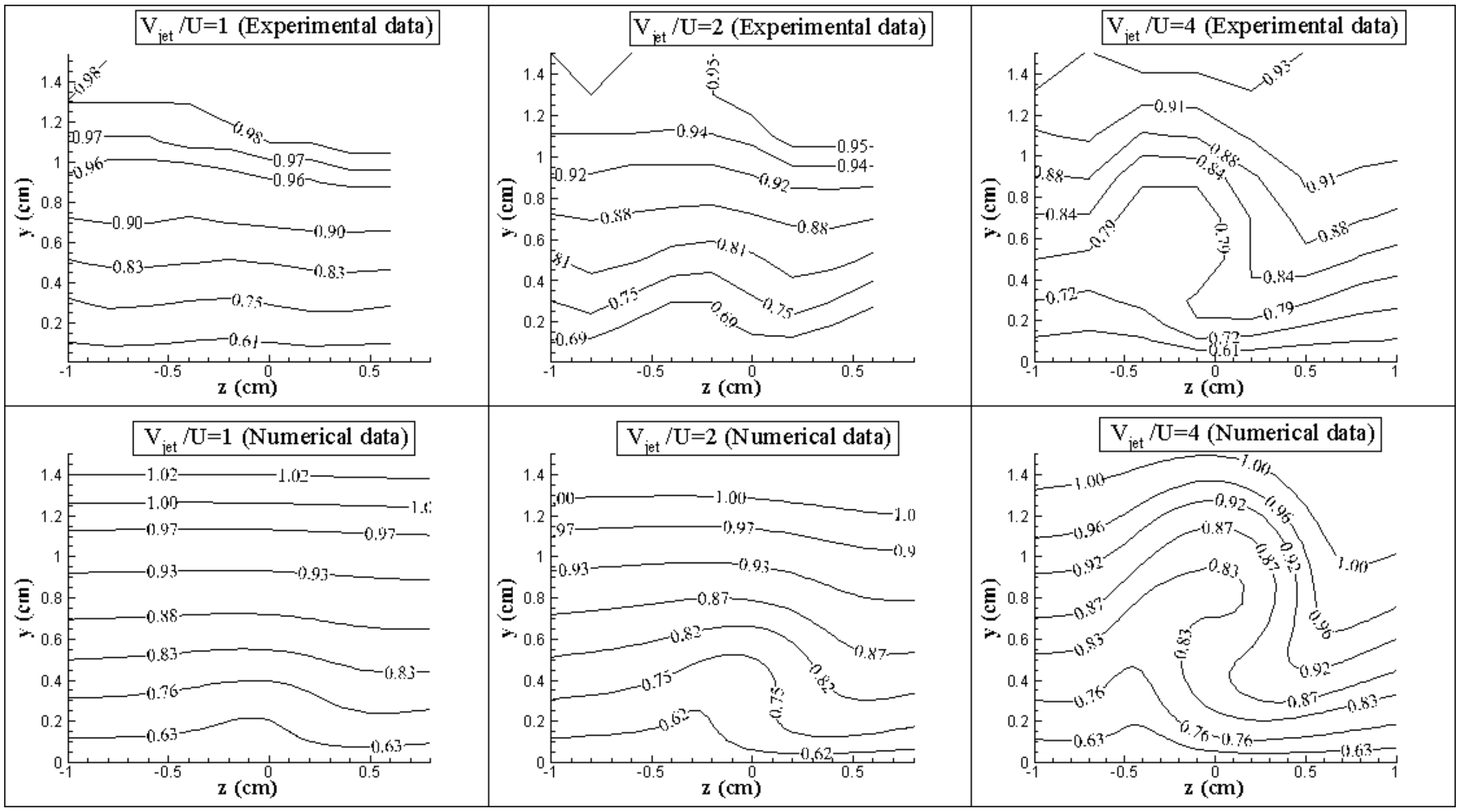

Figure 3: Comparison of the numerical and experimental [25] contour plots of $V_{jet}/U$ for the case with a jet and VRs = 1, 2, and 4.

What follows presents the grid-independence test for the case with a jet. Fig. 4 shows the computational domain for this case. The figure depicts the considered geometry for which a hexahedral (HEX) mesh is used for grid generation. The mesh is finer in regions with more important and intensified flow variations, such as near the VGJs, the lower, and the inlet boundaries of the domain.

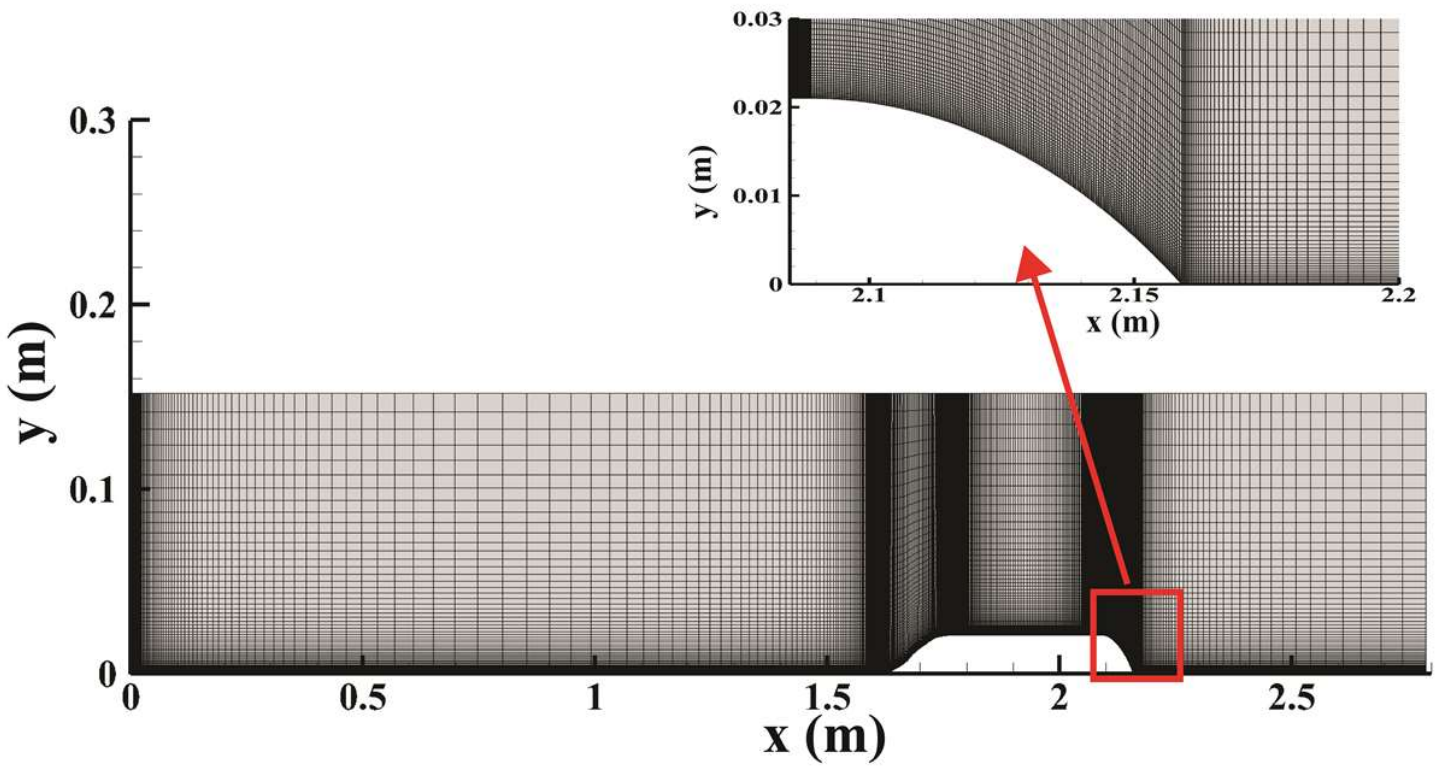

Figure 4: Graphical representation of the mesh grid used for the computational domain.

To carry out the test, the results were obtained for three mesh types, namely (with the number of its cells) coarse (2200000), normal (6400000), and fine (16250000), as shown in Fig. 5. The number of cells was increased along the three directions in each step, especially in regions closer to the jet and the ramp. The figure shows the velocity and Reynolds stress results at three cross-sections, namely before $(x' = -2)$, at the beginning $(x' = 0)$, and at the end of the ramp $(x' = 1)$, where $x' = x/L_1$ is the longitudinal distance from above the ramp. It is observed that all three mesh grids yielded relatively acceptable results. However, the results from the coarse mesh differed from those from the other two, especially in velocity along the $y$-direction and the Reynolds stress $u'v'$. Therefore, the normal mesh was used in further analyses for computational efficiency rather than the fine one.

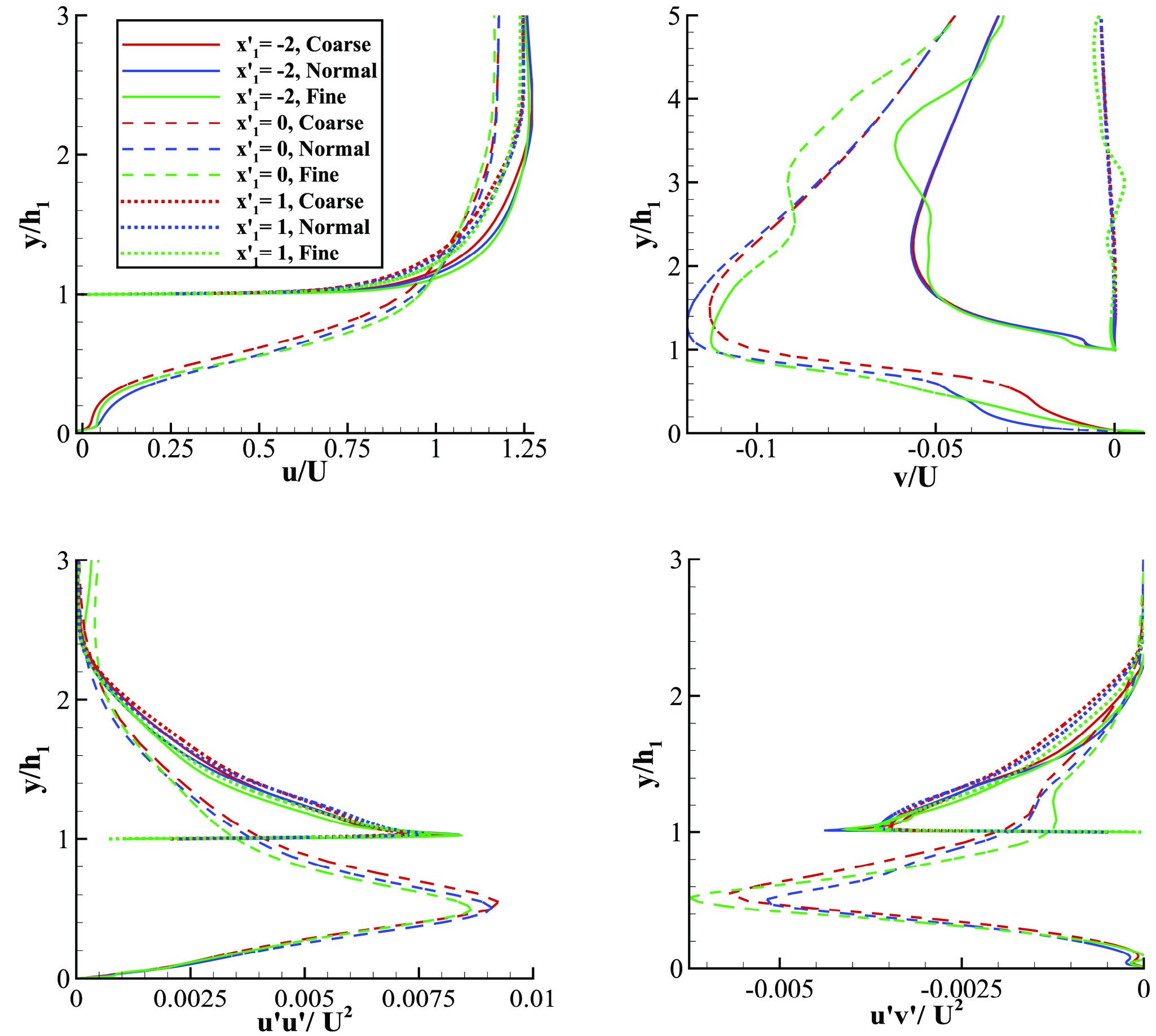

Figure 5: Comparison of the velocity components *u* and *v* and Reynolds stresses u'u' and u'v' for various mesh densities.

## 3. *Results and Discussion*

This section discusses the effects of the fluid jet on the flow separation region, velocities, Reynolds stresses, vorticity, pressure, and streamlines. The analyses are presented for various flow cross-sections separately with the jet at ($L_x/L_l$: Jet 1 = 0.0214, Jet 2 = 0.286, and Jet 3 = 1.357). In what follows, the effects of a single jet on the separation bubble are first presented. The effects of the three-jet array on the flow separation region are discussed next.

### *3.1. Results for a Single Microjet*

It is observed from Fig. 6 (Cross-section 1: *x'* = 0) that the maximum value of *u* occurred near the surface when the jet was placed nearest to the ramp. As the jet was moved farther, the maximum *u* decreased and occurred farther from the surface. As the jet moved yet farther, the flow velocity peak vanished, leading to a more laminar-like flow regime (similar to that at the same cross-section in the case without a jet). The same trend was observed for *v* and *w*; the jet had a weaker influence on the flow near the boundaries as it was moved farther. The velocity contours also confirm that the maximum vorticity experienced a tenfold difference as the jet was moved farther away. For Jet 1, the two initial vortices persisted without merging, while for Jet 2, only one vortex was present, and for Jet 3, a weakened vortex reached the top of the ramp. The effects of changing velocity parameters are revealed in the stress contours. For Jet 1, the stress effects were confined to the boundary layer, and as the jet was moved farther, these effects also gradually moved farther away from the bottom boundary layer. It is also observed that the impact of microjet on *u'w'* were almost half of *u'v'*, while these two stresses maintain similar values in the case of the other jets. The impacts of the vorticities also appear in the pressure contour, where the gradual diminishing of the pressure gradient is evident. In addition, the general flow effects are observed to have moved leftward (to the jet's lateral position) when the jet was moved backward.

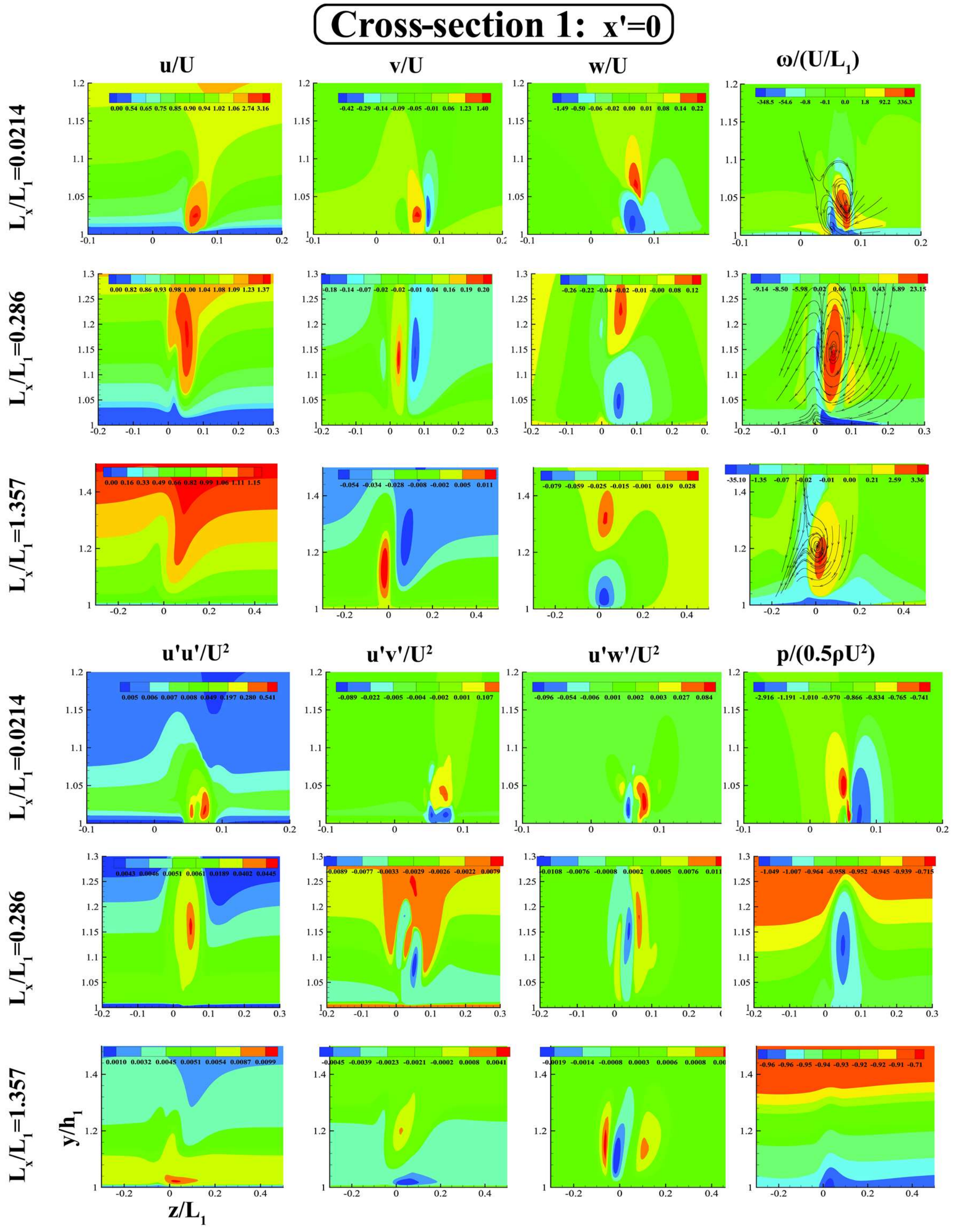

Figure 6: Effects of a single jet on the flow velocity components, vorticity, Reynolds stresses, and pressure at Cross-section 1 ($x'$ = 0).

It is observed from Fig. 7 (Cross-section 2: $x'$ = 0.25) that no separation occurred anywhere along the flow. The impacts of Jet 1 moved up (positive $y$-direction), whereas those of jets 2 and 3 remained almost at the same height, experiencing a slight drop. The two small initial vortices merged for Jet 1, were weakened for Jet 2, and completely vanished for Jet 3. The jet impacts on Reynolds stresses were also reduced. It is also observed from $v$ and $u'v'$ contours that these parameters rose for Jets 2 and 3 near the surface due to the ramp curvature.

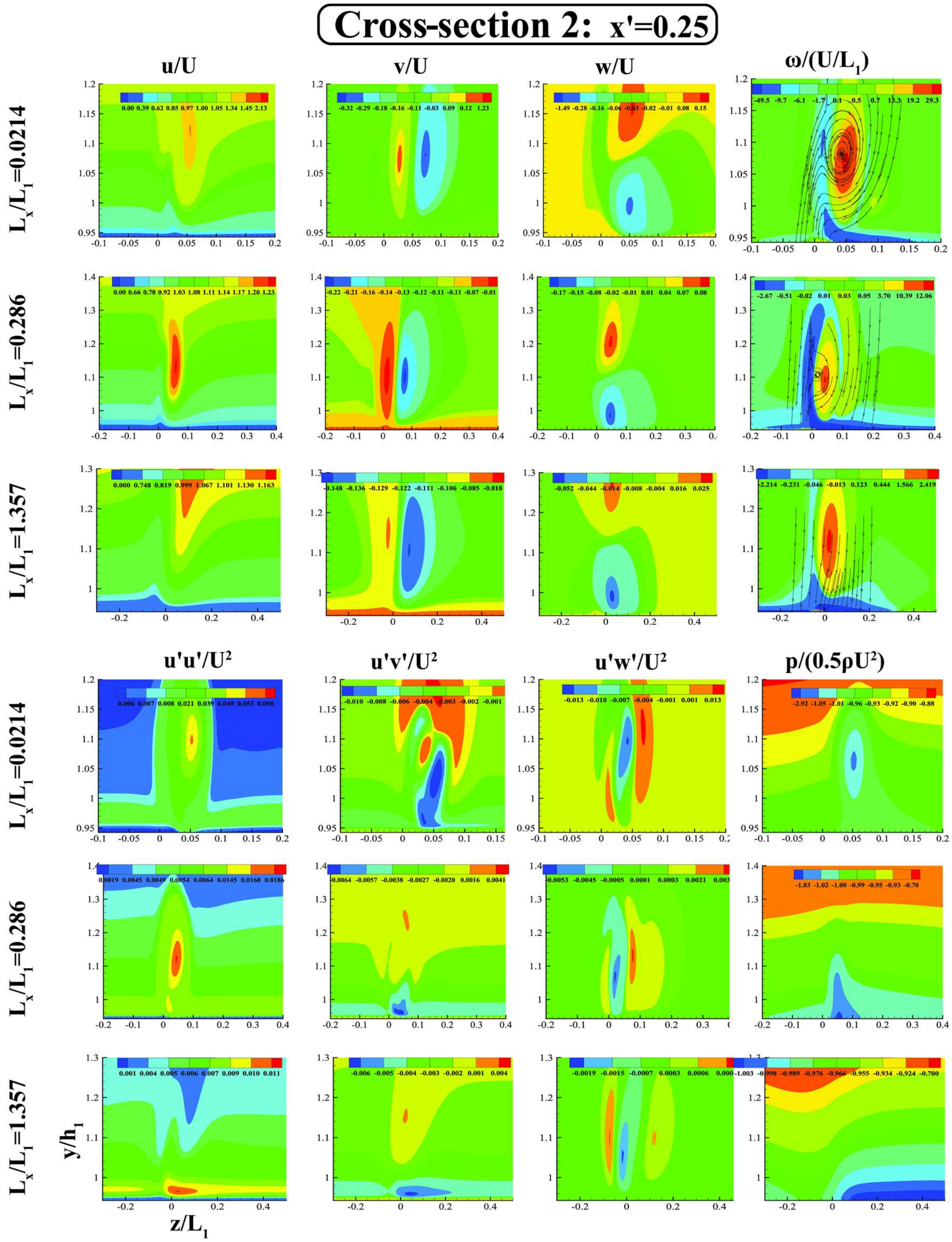

Figure 7: Effects of a single jet on the flow components, vorticity, Reynolds stresses, and pressure at Cross-section 2 ($x'$ = 0.25).

There was still no separation at Cross-section 3 ($x'$ = 0.5), as shown in Fig. 8. The jet impacts were also pulled towards the down boundary for all the jets. The vortices vanished and were no longer visible. The pressure value and gradient were also considerably reduced. It is observed from the figure that Jet 3 was able to hold its impacts to a small extent and affected the boundary layer on the ramp. In fact, from a distance onward, the minor flow variations persisted for a while almost constantly.

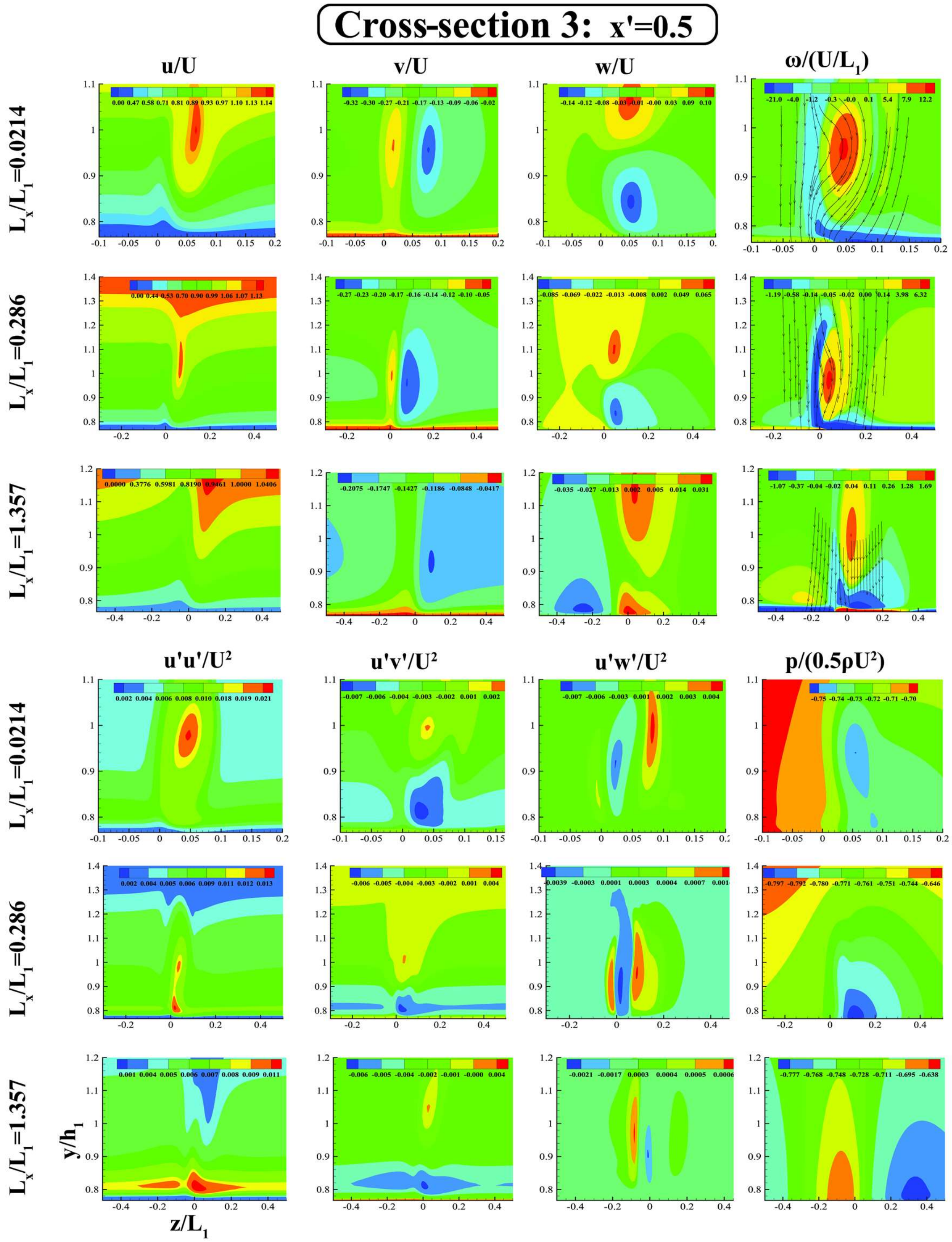

Figure 8: Effects of a single jet on the flow components, vorticity, Reynolds stresses, and pressure at Cross-section 3 ($x'$ = 0.5).

Backflow occurred at Cross-section 4 ($x'$ = 0.75), as indicated by the deep blue in the $u/U$ contour shown in Fig. 9. The constant backflow height is observed in the $u/U$ contour and far-jet regions, higher in the upwash. No backflow is observed in the region affected by the jet. In the downwash region, the backflow height also decreased as long as the jet effects persisted in its trail. As the jet was moved away, the flow height decreased accordingly. This backflow, however, was increased more in the downwash. There was again an insignificant rise in vorticity. The same effects are observed in the distribution of the Reynolds stresses, although with no significant change in value. The pressure also assumed a different distribution despite its lower variations.

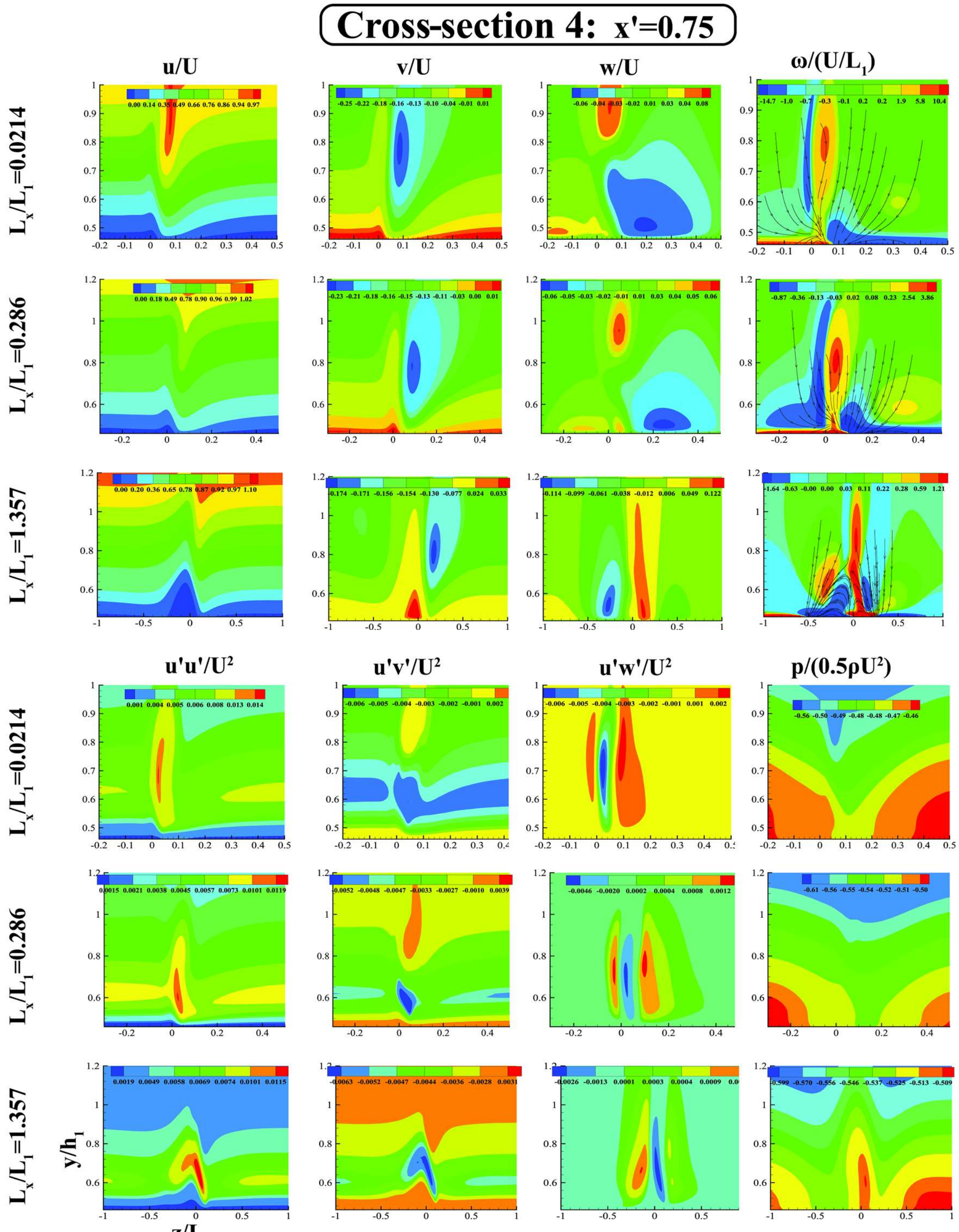

Figure 9: Effects of a single jet on the flow components, vorticity, Reynolds stresses, and pressure at Cross-section 4 ($x'$ = 0.75).

As shown in Fig. 10, the backflow height peaked at Cross-section 5 ($x'$ = 1), where the jets could not eliminate the backflow. The upwash height for Jet 3 was also great in this case. As $v$ and $w$ increased in the upwash and downwash regions due to the flow effects running into the bottom, the resulting pressure drop in those regions gave rise to counter-rotating vortices. These variations also led to new Reynolds stress peaks along the flow.

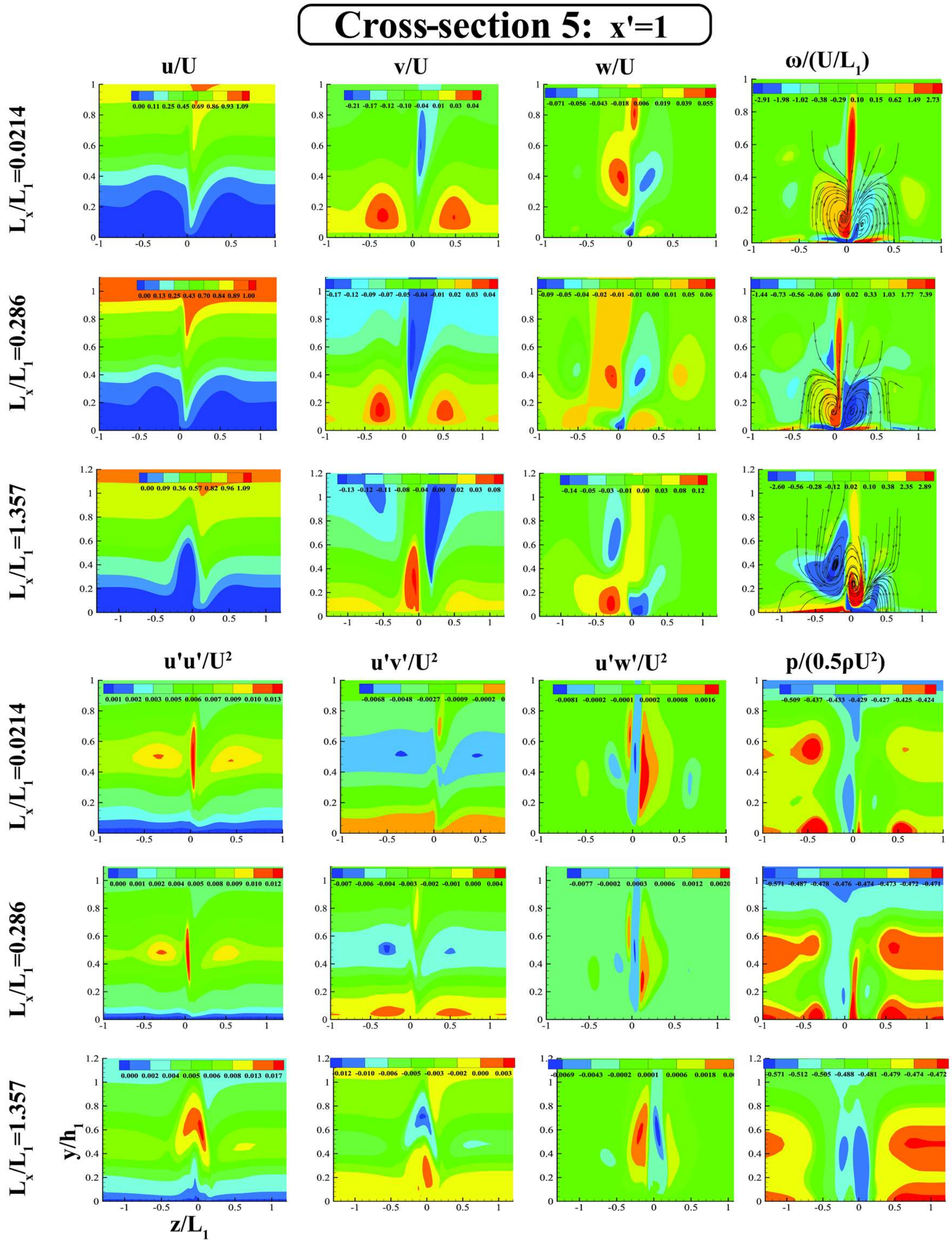

Figure 10: Effects of a single jet on the flow components, vorticity, Reynolds stresses, and pressure at Cross-section 5 ($x'$ = 1).

The vortices gradually separated and moved away from the bottom at Cross-section 6 ($x'$ = 1.086), shown in Fig. 11. The backflow persisted in regions far from those affected by the jet. However, the backflow was eliminated in the regions where the jet's effects were present. These regions were almost the same for the three jets. Nevertheless, there was a severe backflow for Jet 3 in the upwash. The Reynolds stress did not experience a significant change.

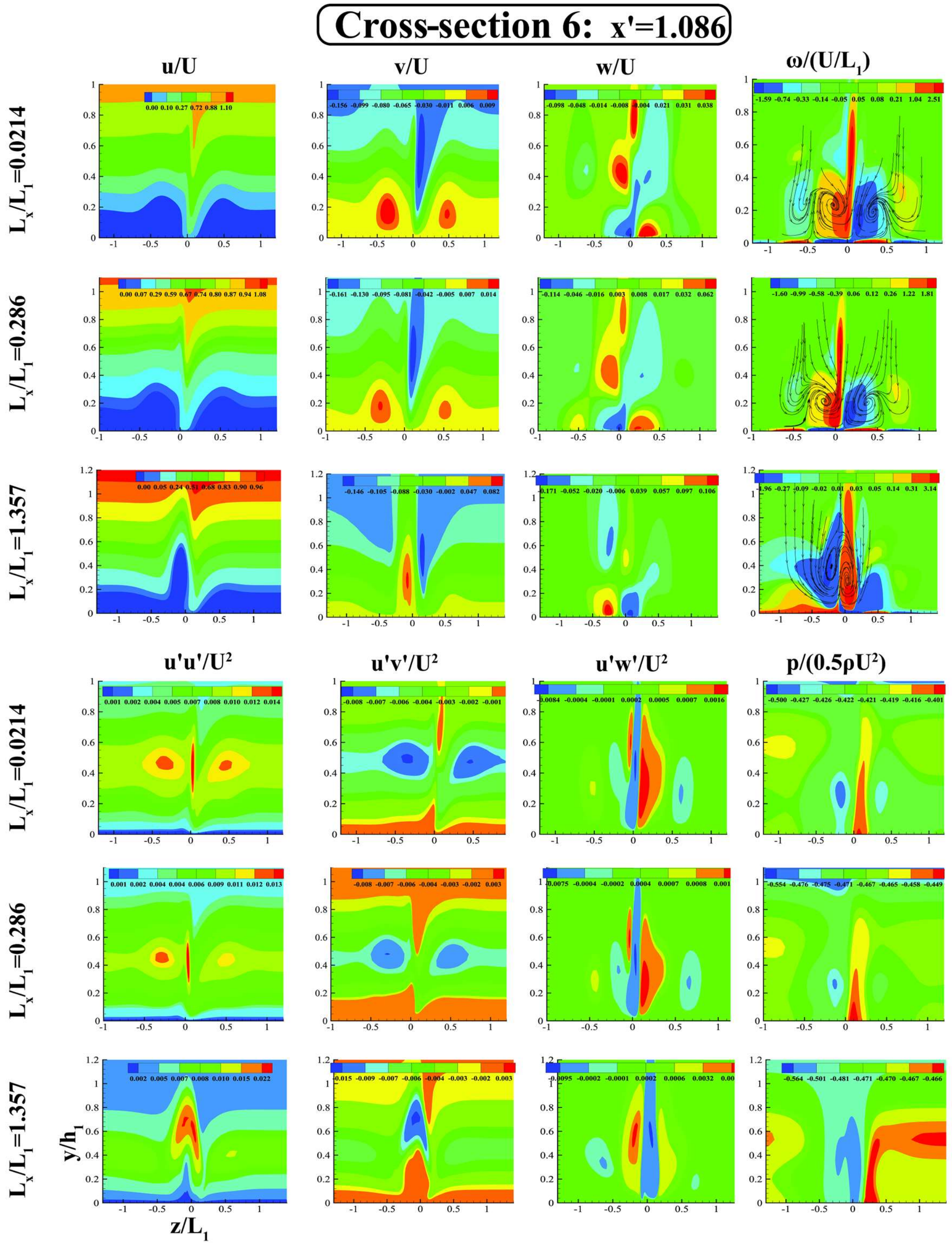

Figure 11: Effects of a single jet on the flow components, vorticity, Reynolds stresses, and pressure at Cross-section 6 ($x'$ = 1.086).

The jet's effects are extended to Cross-section 7 ($x'$ = 1.3), as shown in Fig. 12, although they were generally weakened throughout the cross-section. The vortices for the first two jets vanished entirely and are not depicted. For Jet 3, the vortices merged almost wholly, probably due to this jet's larger upwash and downwash regions. The Reynolds stresses were present with insignificant variations compared with those of the previous section.

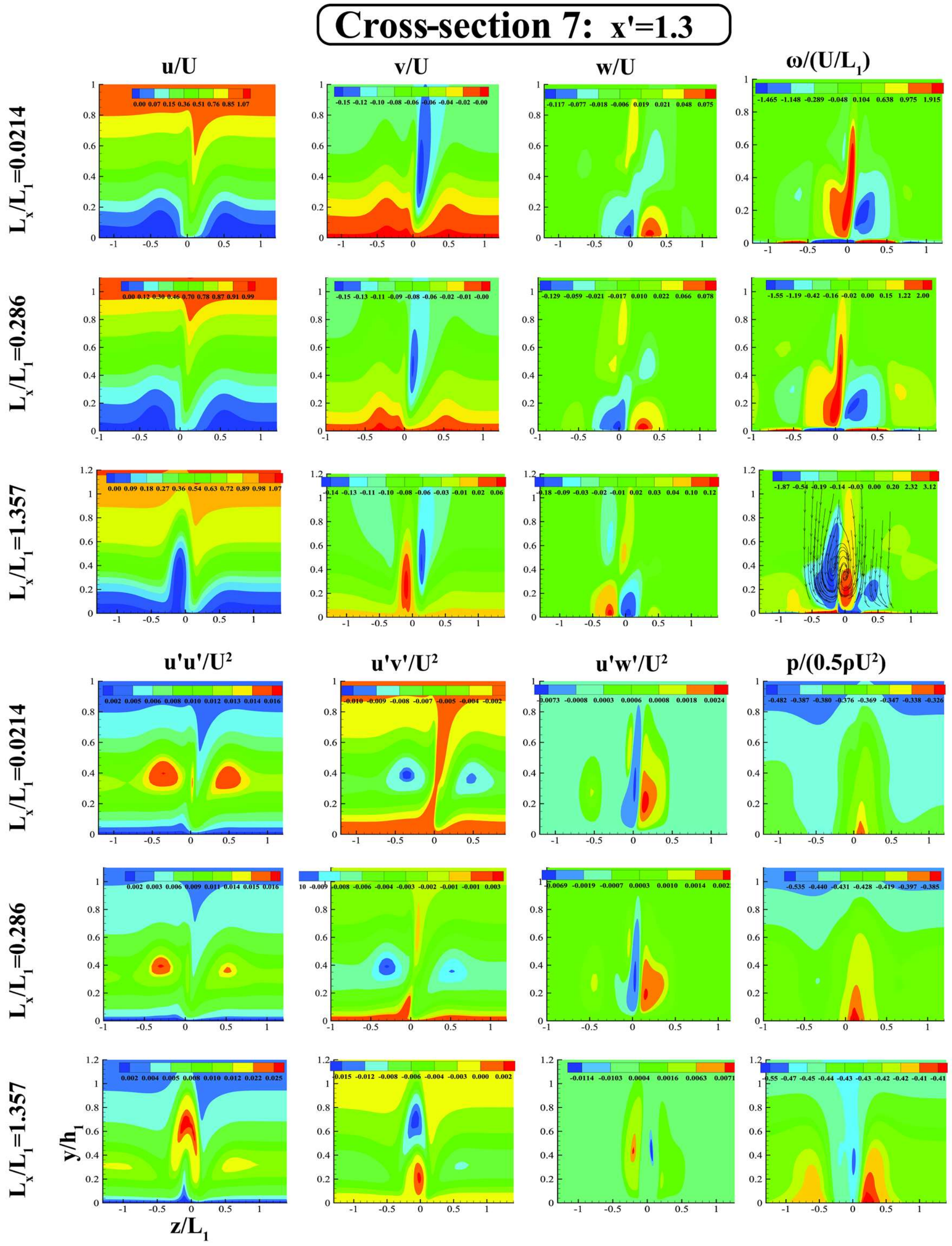

Figure 12: Effects of a single jet on the flow components, vorticity, Reynolds stresses, and pressure at Cross-section 7 ($x'$ = 1.3).

In Cross-section 8 ($x'$ = 1.44) shown in Fig. 13, the backflow is observed only in the upwash and downwash regions. The backflow was eliminated in the intermediate and far regions. Because of turbulence, traces of Reynolds stresses were still present in the flow. The results were similar for the first two jets, with a slight change for Jet 3, which kept the vortices merged.

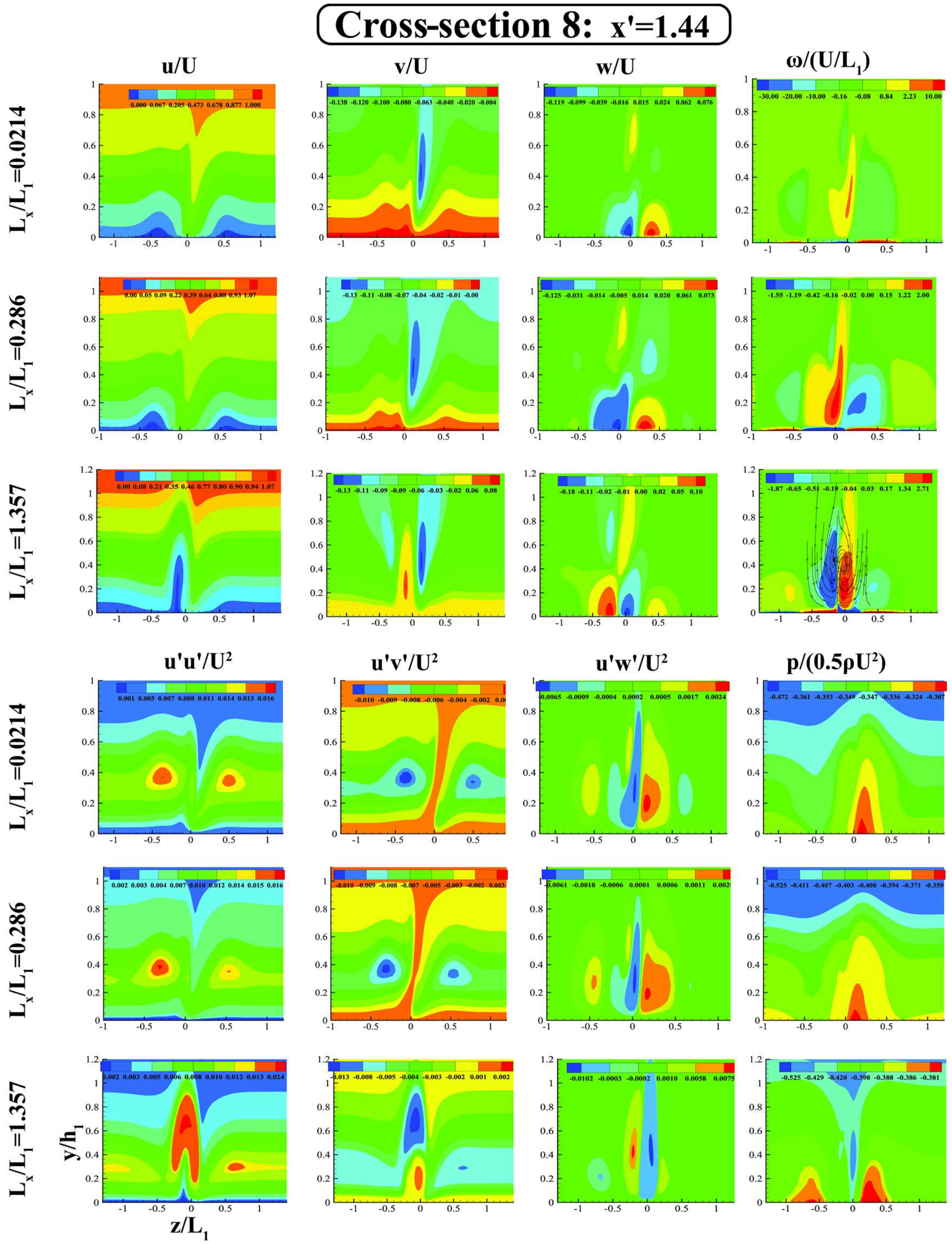

Figure 13: Effects of a single jet on the flow components, vorticity, Reynolds stresses, and pressure at Cross-section 8 ($x'$ = 1.44).

Regarding the obtained contours, the following figure illustrates the graphs of the *z*- and *y*-positions of the maximum backflow and the separation region eliminated by the jet. According to Fig. 14, no difference is observed between the results for Jets 1 and 2 in terms of the maximum backflow's lateral (*z*) position. This position gradually moved leftward for both of these jets until it almost settled at a certain point. For Jet 3, this position was closer to $z = 0$ and nearly stationary. The backflow amplitude increased as the jet was moved back. The flow separation in the upwash regions of Jets 3 and 2 was about 1.2 and 2 times that of Jet 1, respectively.

Moreover, the three jets performed similarly in separation bubble control along the lateral ($z$) direction. The backflow region grew more extensive along the flow. In fact, for all the three jets, this (dimensionless $\Delta z$ separation region) was about 0.2 and 0.6 at the beginning and end of the effective region, respectively, with Jets 1 and 2 performing better. It can be concluded that Jets 1 and 2 performed similarly but outperformed Jet 3 in creating a region for keeping the flow from separation.

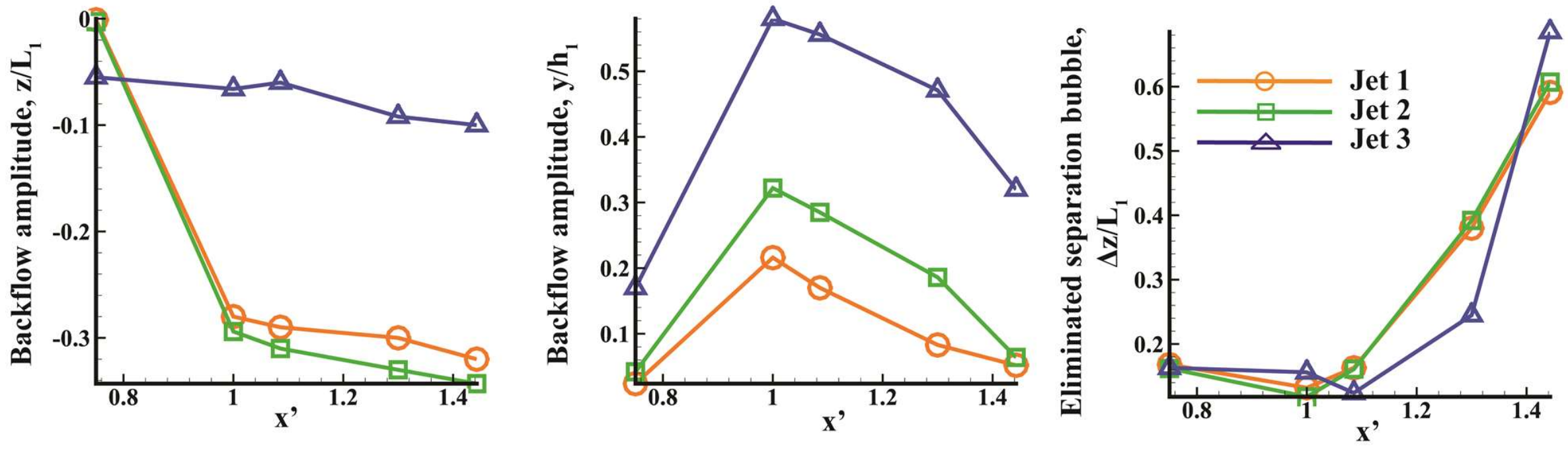

Figure 14: $z$- and $y$-positions of the maximum backflow and the separation region eliminated by the three jets across the computational domain.

Fig. 15 provides a velocity graph along the $x$-direction over the ramp geometry for various cross-sections based on the previously presented contours and graphs. The results are given for the cases with/without a jet at positions 1 to 3. The graphs were analyzed for the center region (where separation was delayed) and the upwash region (where the largest separation occurred). According to the results for the center shown in Fig. 15, using all three jets delayed flow separation in all the regions, with a stronger effect from Jet 1 (which generated a higher velocity along the $x$-direction than in the other cases). At points farther from the bottom level and also away from the vortex effects downstream the ramp, the flow, especially for Jet 3, is more likely to experience separation. In this region, Jet 1 outperformed the other two. According to the results for the upwash shown in Fig. 15, using the jets aggravated the separation. This was particularly more severe for Jet 3. It performed worse than the other two in terms of both flow separation strength (negative velocity magnitude) and the backflow amplitude. Jets 1 and 2 performed highly similarly in this region.

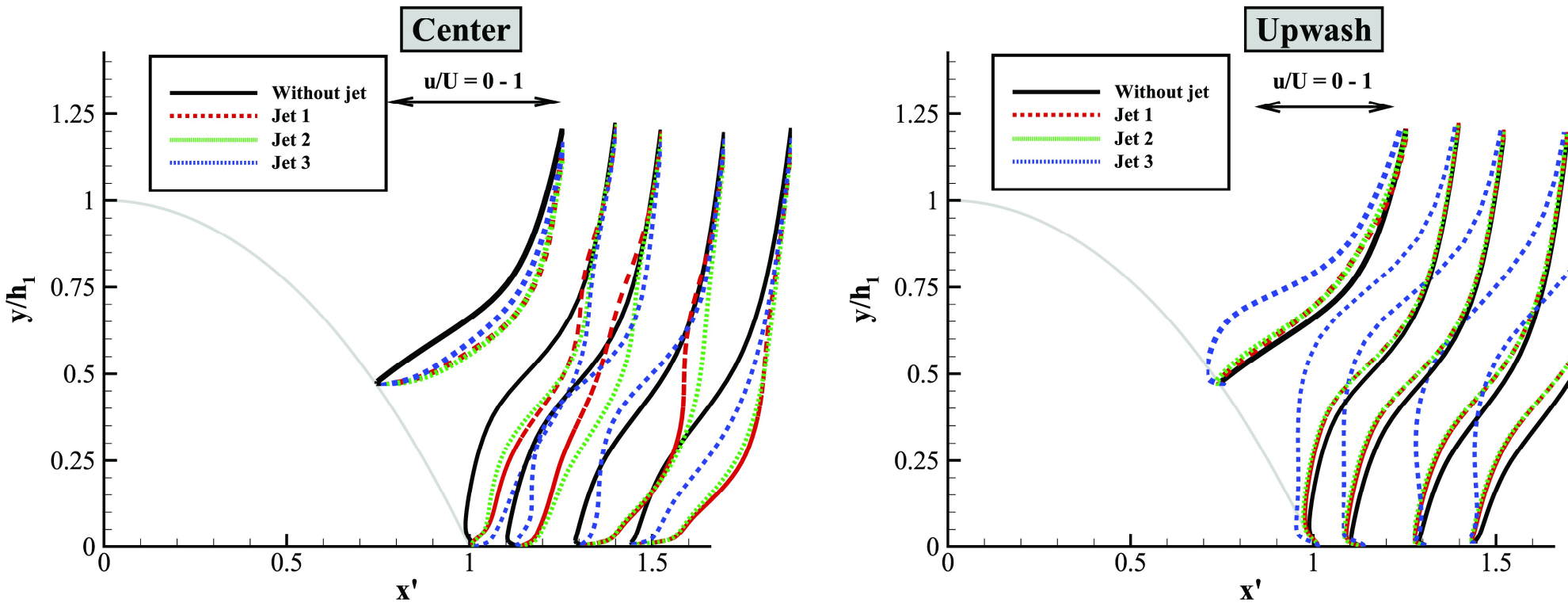

Figure 15: Comparison of velocity profiles along the *x*-direction for the cases without and with jet across the computational domain.

### *3.2. Results for an Array of Microjets*

This section examines the effects of an array of microjets on post-ramp flow separation. To this end, the results are investigated for the case of employing no jet and the case of introducing an array of jets (three jets) above the ramp and changing their longitudinal distance from it and their cross distance from one another. Then, we will examine the entrainment zone. It is expected that we observe, in practice, the ability of a set of jets to control the flow separation. To do so, using these jets must push the flow separation zone as much as possible. Fig. 16 depicts the schematics of flow separation control by microjets array.

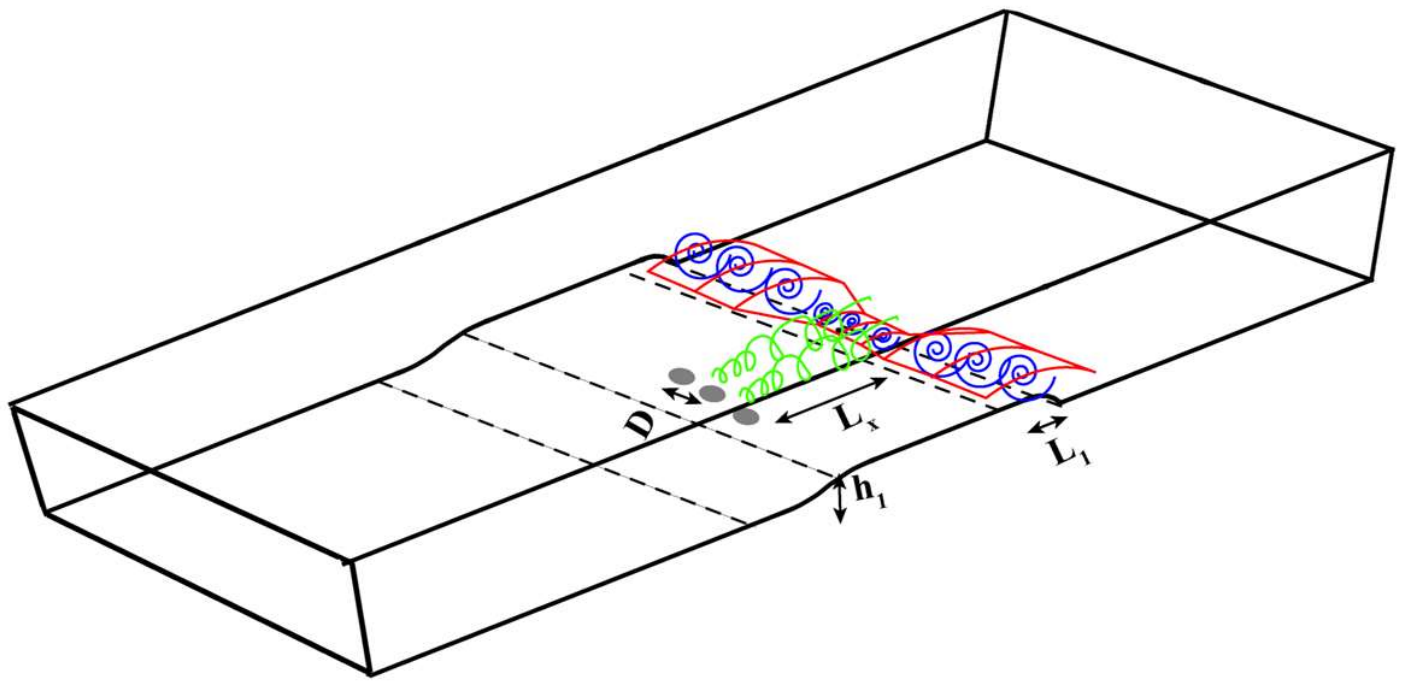

Figure 16: Schematic representation of flow separation control by an array of three microjets.

The effects of an array of three microjets were analyzed by placing the jets at *d*/*D* = 10, 15, and 30 (*d* = space between a pair of microjets, *D* = diameter of the microjet) intervals from each other and also at the same position as that of Jet 1. Fig. 17 presents the velocity along the *x*-direction and flow pattern at different cross-sections.

For *d*/*D* = 10, no significant vortex interaction is observed, and hence, the intervals between the jets can be increased for optimization. It is also observed that the three generated vortices moved similarly to those in the previous case and dissipated at a distance into the flow. After that, as the vortex effects ran into the bottom, two larger vortices (compared to the case of single Jet 1) were formed and dissipated at a distance into the flow. The effects of the jets on the separation region were analyzed by considering Cross-section 4, where the upwash and downwash regions of the internal jets vanished, leaving no trace along the flow (deep blue indicates the separation bubble). However, three-jet and single-jet cases were similar in the upwash and downwash regions with the same backflow height as before. It is observed that using three jets led to a separation bubble region about 2.1 times that of the single-jet case (this value is predicted to rise to 2.6 through optimization). Moving forward through the flow, the vortex effects expanded, and the impacts of the middle jet through the flow separation were merged into those of the other two vortices. However, at Cross-section 7, the effects of three jets remained about 2.1 times those of a single jet. Therefore, the three jets evidently increased the dissipated region width of the separation bubble by almost two times that in the single-jet case. Furthermore, since the jets did not have enough time to merge, their strength remained the same, and therefore, this region was as long and high as in the previous case.

For optimization purposes and prevention of flow separation in a larger region using the same number of jets, the results were also obtained for $d/D$ = 15 and 30. It was observed that $d/D$ = 10 and 15 yielded similar results and the same number of generated vortices. There was no upwash within the region of flow separation prevention. Therefore, the jets can be placed at $d/D$ = 15 intervals for cost-saving. The region where flow separation was prevented was 2.75 times more than in the case with a single jet. As the interval between the jets was increased, and the jets were placed at $d/D$ = 30, upwash effects were observed within the region of flow separation prevention. It was concluded that flow separation occurred in this region and that $d/D$ = 30 is not a suitable interval.

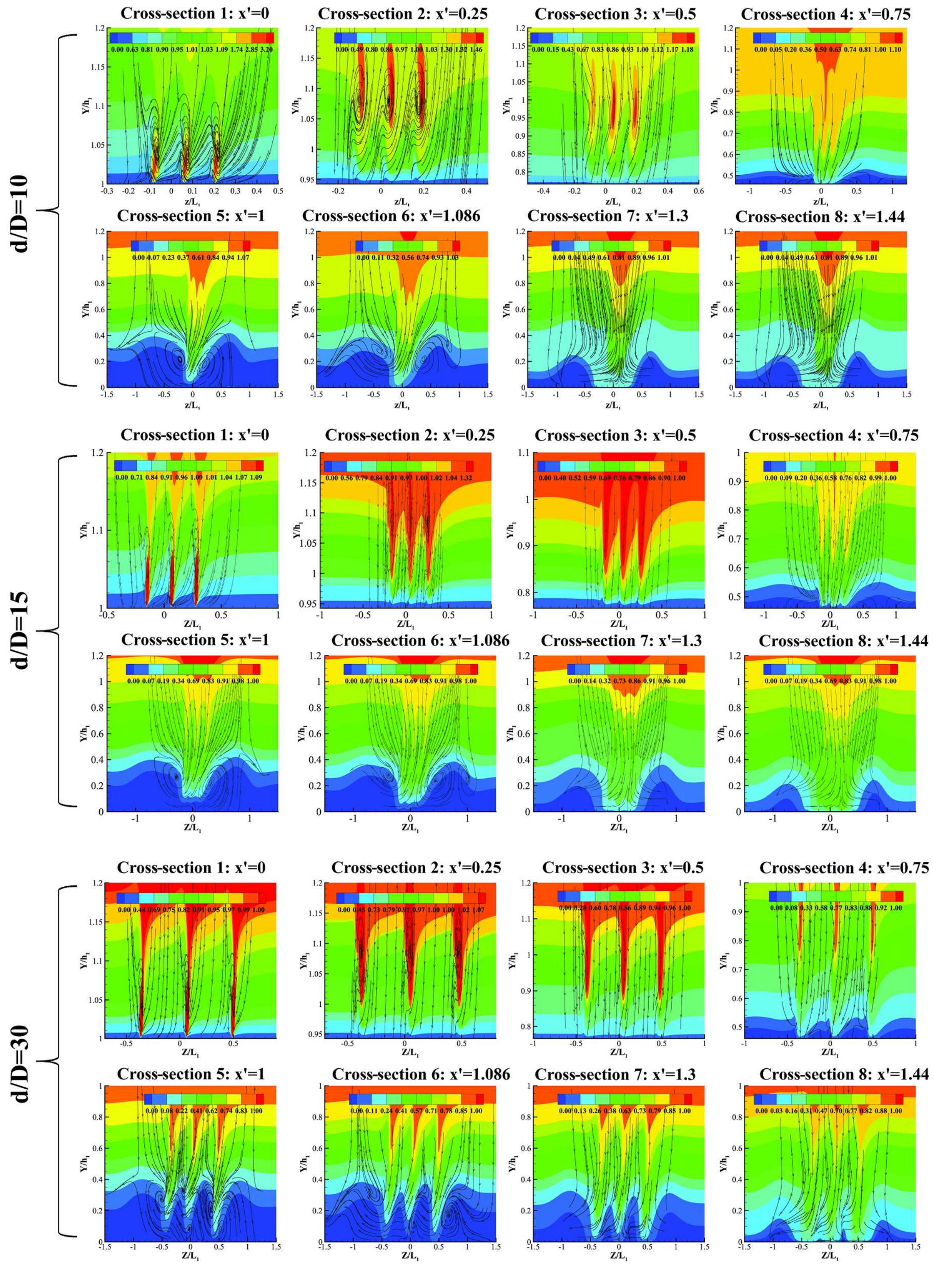

Figure 17: Velocity contours and flow pattern for the case of a three-jet array for $d/D$ = 10, 15, and 30 at different cross-sections.

Velocity contours and flow lines corresponding to the changes in the longitudinal coordinate of jets and their distance from one another in the $x$-$y$ plane are shown in Fig. 18 and 19 for the section in which $z/L_1$ = 0.07143 (the cross coordinate of the middle jet). In order to do so, the results for the case of using no jet in which ($L_x/L_1$ =0.357, $d/D$ =15), ($L_x/L_1$ =1.43, $d/D$ = 15), ($L_x/L_1$ = 0.0143, $d/D$ = 10), ($L_x/L_1$ =

0.0143, $d/D$ = 15) and ($L_x/L_1$ = 0.0143, $d/D$ = 30) have been examined. According to the figure, it is clear that all the arrays of jets have decreased the length and height of the entrainment zone (the separation zone is shown in strong blue). It is also seen that the separation zone has not changed significantly in the $x$-$y$ plane by changing the longitudinal coordinate of jets relative to the top of the ramp. Furthermore, decreasing the distance of the jets from one another, the vortices combined and their strength increased consequently, and the separation zone in the $x$-$y$ plane declined. Again, the flow return is evident in this zone regarding the flow lines. With the introduction of jets, the center of rotation of the flow lines has gotten closer to the corner of the ramp, and hence the rotation zone and the return zone have declined.

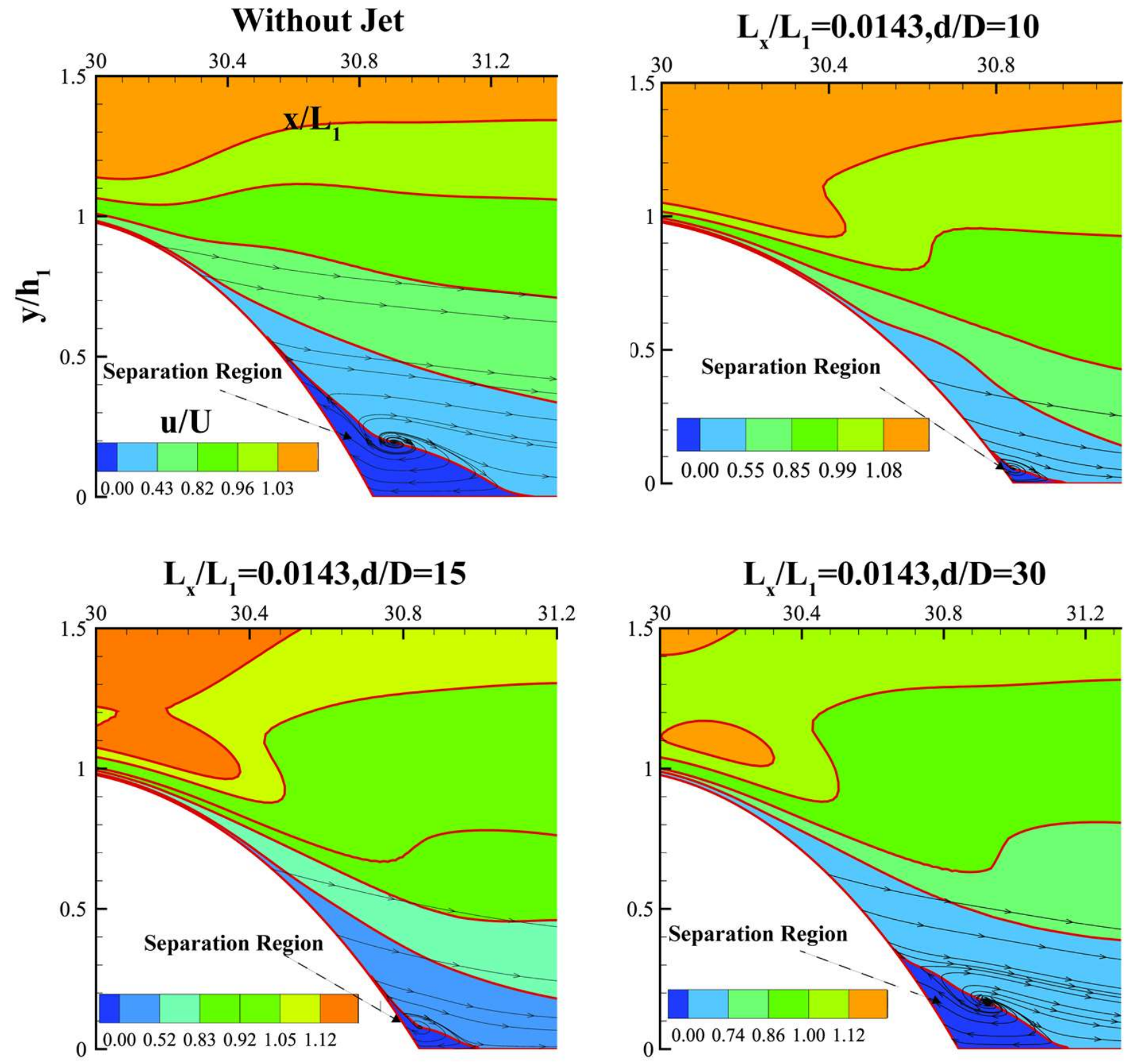

Figure 18: Velocity contours in the longitudinal plane as a function of the cross distance of jets from one another.

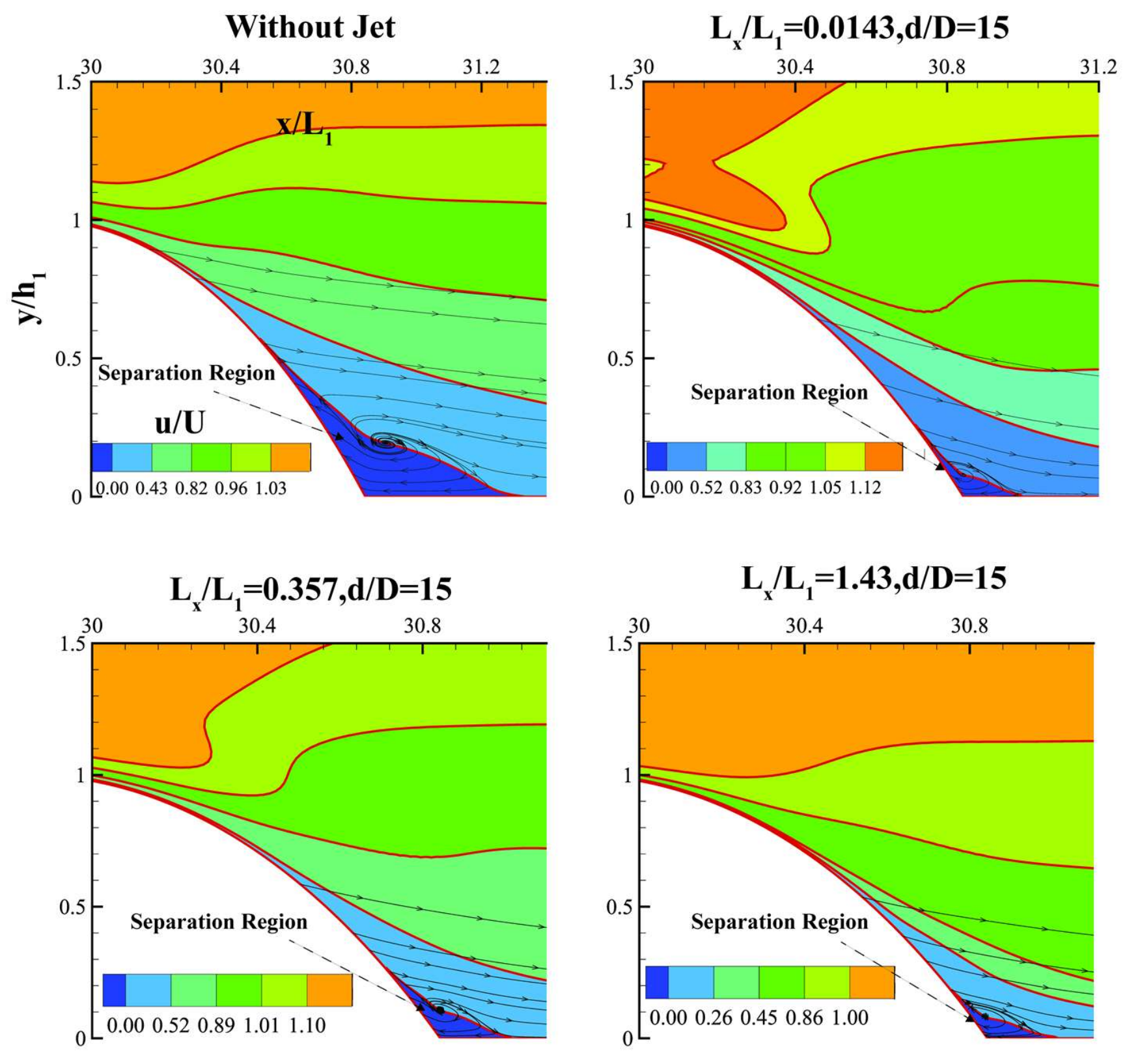

Figure 19: Velocity contours in the longitudinal plane as a function of the longitudinal distance of jets from one another.

The velocity contours and the flow lines corresponding to changes in the longitudinal coordinate of jets and their cross distance from one another in the *y-z* plane are presented in Fig. 20 and 21 for the section for which $x/L_1$ = 30.843 (the position of the lower corner of the ramp). In order to do so, the results for the case of using no jet in which ($L_x/L_1$ = 0.357, $d/D$ = 15), ($L_x/L_1$ = 1.43, $d/D$ = 15), ($L_x/L_1$ = 0.0143, $d/D$ = 10), ($L_x/L_1$ = 0.0143, $d/D$ = 15) and ($L_x/L_1$ = 0.0143, $d/D$ = 30) were compared. According to the figure, it is clear that the introduction of an array of jets has decreased the flow separation zone to some extent. Moreover, with the increase in the distance of the jets array from the ramp, the strength of vortices has declined a little, and it has led to a slight decrease in the prevention of the flow lines from separation. It is also clear that the displacement of the jets to the back has caused the separation to get an extreme rise in the last portion of the jets' effect zone.

Additionally, it is clear that the mentioned prevented zone has become larger with the increase in the distance of the jets from one another. Nevertheless, it is seen that the jets have worked independently by an excessive increase of these distances, and so the separation zone has increased in middle zones. Considering the flow lines, it is also seen that the jets work independently, and each creates its vortex pair in this section when increasing their distance from one another.

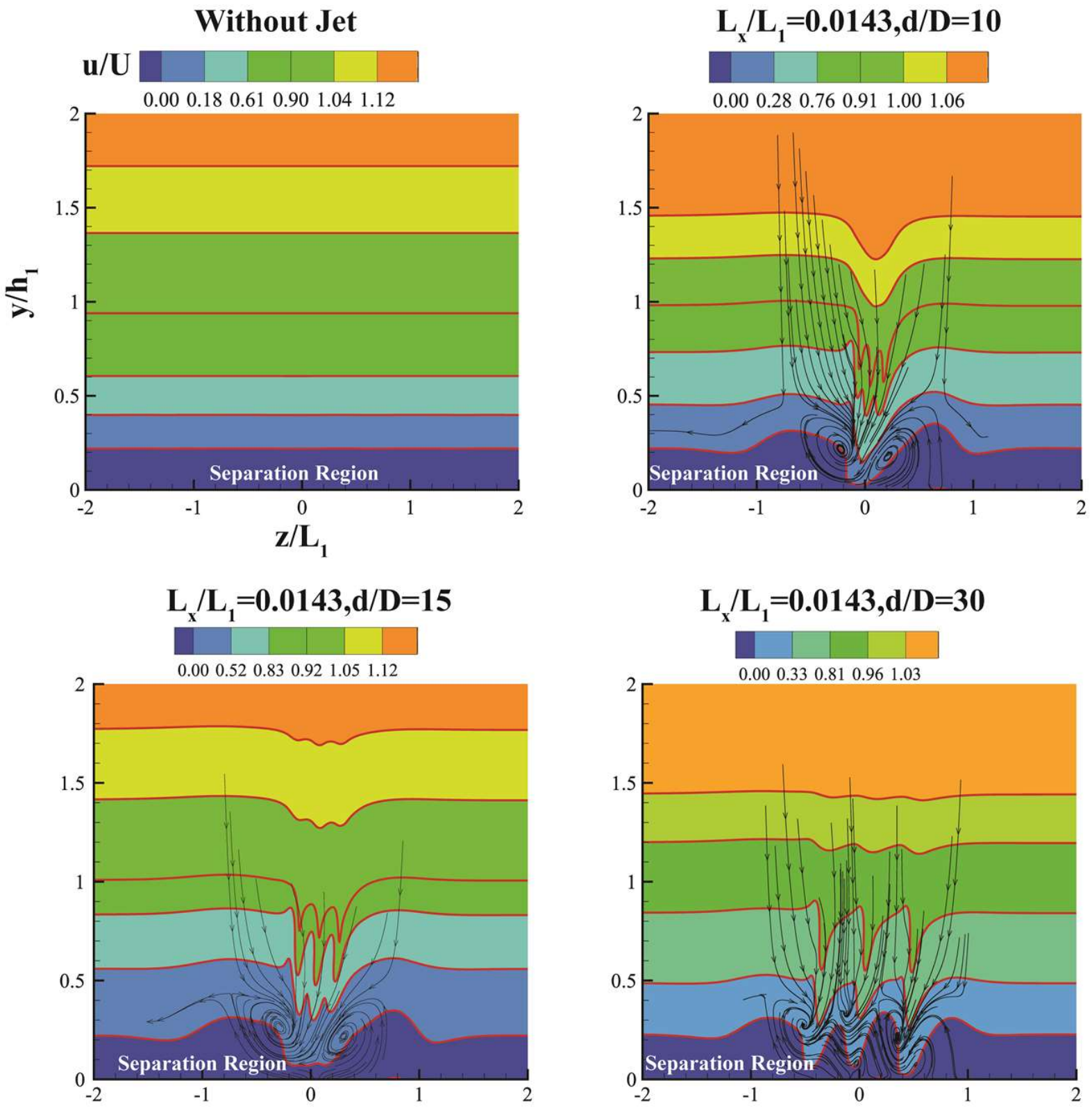

Figure 20: Velocity contours in the cross-plane as a function of the cross distance of jets from one another.

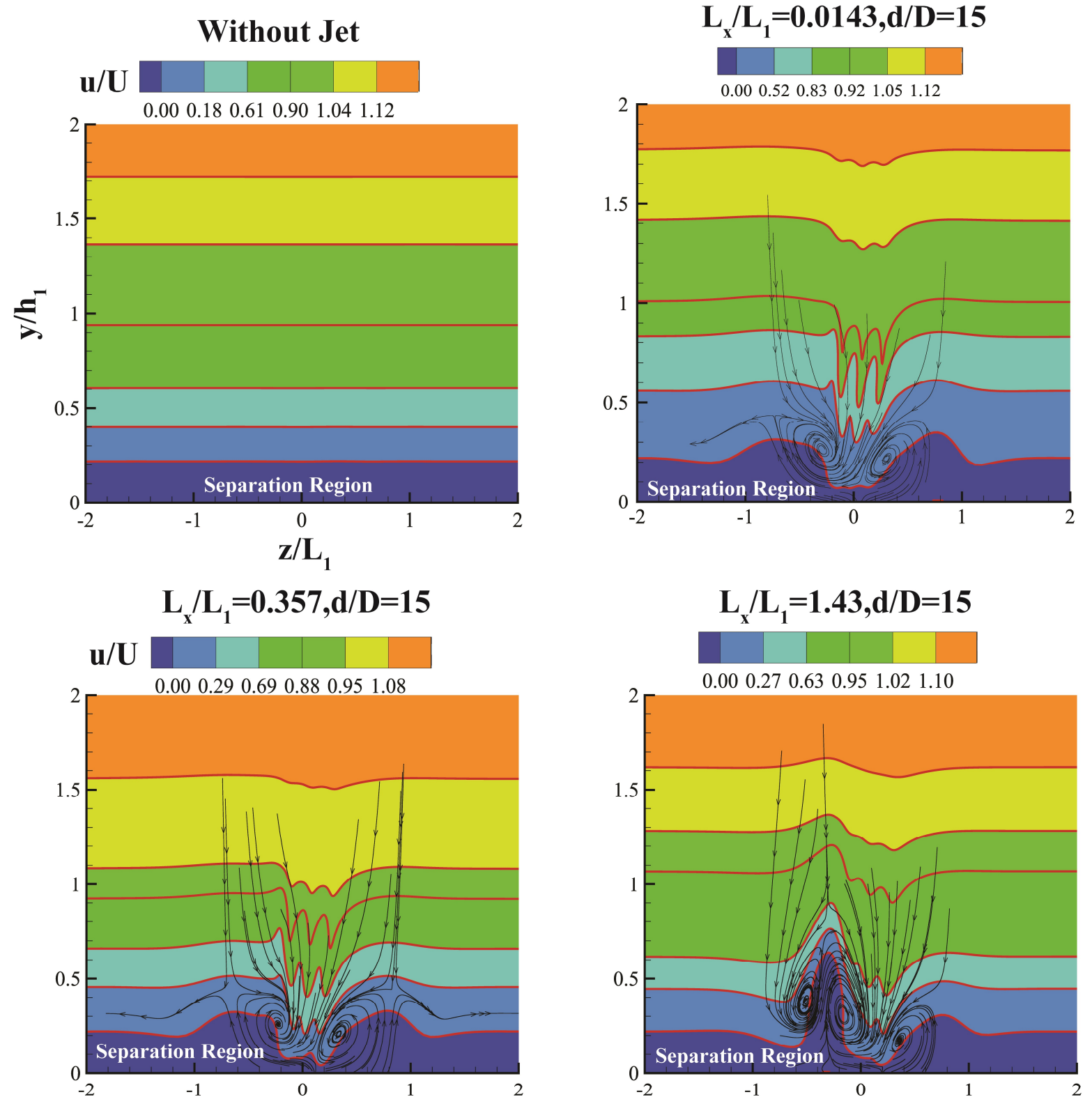

Figure 21: Velocity contours in the cross-plane as a function of the longitudinal distance of jets from one another.

Based on the obtained results and previous discussions, we made a numerical analysis of the critical parameters of these results summarized in Tab. 1. According to the table, the longitudinal coordinate of the detachment point moves ahead with the introduction of jets. In this situation, the case with $L_x/L_1$ = 0.0143 and $d/D$ = 10 has had the best performance causing the detachment point to move 0.25 ahead. This array has effectively decreased the flow separation height by about 0.433. Considering the results for the reattachment point, the best performance belongs again to $L_x/L_1$ = 0.0143 and $d/D$ = 10.

Hence, this array of jets has had the best performance by reducing the length of the separation zone from 0.75 to 0.16 (equal to 78%). It is also clear that changing the longitudinal coordinate of jets has not had a considerable effect on this matter. In addition, the arrangement with $L_x/L_1$ = 0.0143 and $d/D$ = 15 has had the best performance with a 71% reduction in the length of the separation zone in the *x*-direction. Considering the flow separation width results in the *y-z* plane, it is clear that the arrangement with $L_x/L_1$ = 0.0143 and $d/D$ = 15 has had the best performance. Moreover, the separation zone width has declined due to displacing the jets to the back and reducing their relative distance. By comparing the results for ($L_x/L_1$ = 0.0143 and $d/D$ = 15) and ($L_x/L_1$ = 0.0143 and $d/D$ = 10), it can be observed that the separation zone has declined about 0.2 by the 5-unit increase in distance of jets from one another. According to the results for minimum reduction in the height of the separation zone, it is found that the jets' arrays have had almost similar performances, and there has been no considerable change in the results. Altogether, $L_x/L_1$ = 0.0143, $d/D$ = 10 has had a better performance in this case, too.

Table 1: The detachment and reattachment points in longitudinal and cross-planes for different arrangements of jets.

| | | **Without jet** | **$L_x/L_1$ = 0.0143 $d/D$ = 10** | **$L_x/L_1$ = 0.0143 $d/D$ = 15** | **$L_x/L_1$ = 0.0143 $d/D$ = 30** | **$L_x/L_1$ = 0.357 $d/D$ = 15** | **$L_x/L_1$ = 1.43 $d/D$ = 15** |
|---|---|---|---|---|---|---|---|
| **Separation point** | **$x/L_1$** | 30.54 | 30.79 | 30.76 | 30.69 | 30.74 | 30.76 |
| **Reattachment point** | **$x/L_1$** | 31.29 | 30.95 | 30.98 | 31.137 | 30.99 | 30.995 |
| **Length of separation region** | **$dx/h_1$** | 0.75 | 0.16 | 0.22 | 0.447 | 0.25 | 0.235 |
| **Width of separation region** | **$dz/L_1$** | - | 0.5044 | 0.706 | - | 0.657 | 0.596 |

According to our findings, it can be said that the best performance belongs to the jets' array in which $L_x/L_1$ = 0.0143 and $d/D$ = 15. Although $L_x/L_1$ = 0.0143 and $d/D$ = 10 could have a better performance of about 7% in the longitudinal direction, the separation width reduction, in this case, is 29% less than that in the earlier one. Therefore, the array for which $L_x/L_1$ = 0.0143, $d/D$ = 15 can be considered an optimum for jets' longitudinal and cross positions.

Velocity contours for various arrangements of jets are shown in Fig. 22. Within the figure, velocity contours for the section in which $z/h_1$ = 0.0714 as well as flow lines for the case without jet and the cases of positioning the jets in the arrangements with $L_x/L_1$ = 0.0143 and $d/D$ = 10, $L_x/L_1$ = 0.0143 and $d/D$ = 15, and $L_x/L_1$ = 0.0143 and $d/D$ = 30 are shown. Also, the separation zone is depicted in blue color. According to the figure, it is clear that the jets have disturbed the velocity profile in the boundary layer and have caused the velocity to increase near the boundary. In addition, employing jets has reduced the separation zone. The reduction of the separation zone depends on the positions of the jets. By comparing the results, it is observed that the jets for which $d/D$ = 10 and $d/D$ = 15 have reduced the separation zone well. It is also realized that $d/D$ = 15 has decreased the separation zone width.

Moreover, using the jets in upwash and downwash sections has led to negative performance in controlling the flow separation. Nevertheless, we can transfer such zones far away by increasing the number of jets in engineering practices. This issue can be realized by comparing the results we found here and that of a single jet. Based on these results, we can enhance the removal of separation zone and transfer undesirable zones far away by using jet arrays. Additionally, it can be realized by examining the flow lines that the jets managed to remove the flow rotation at the lower corner of the ramp. Despite the mentioned facts, the case of $d/D$ = 30 did not have appropriate performance.

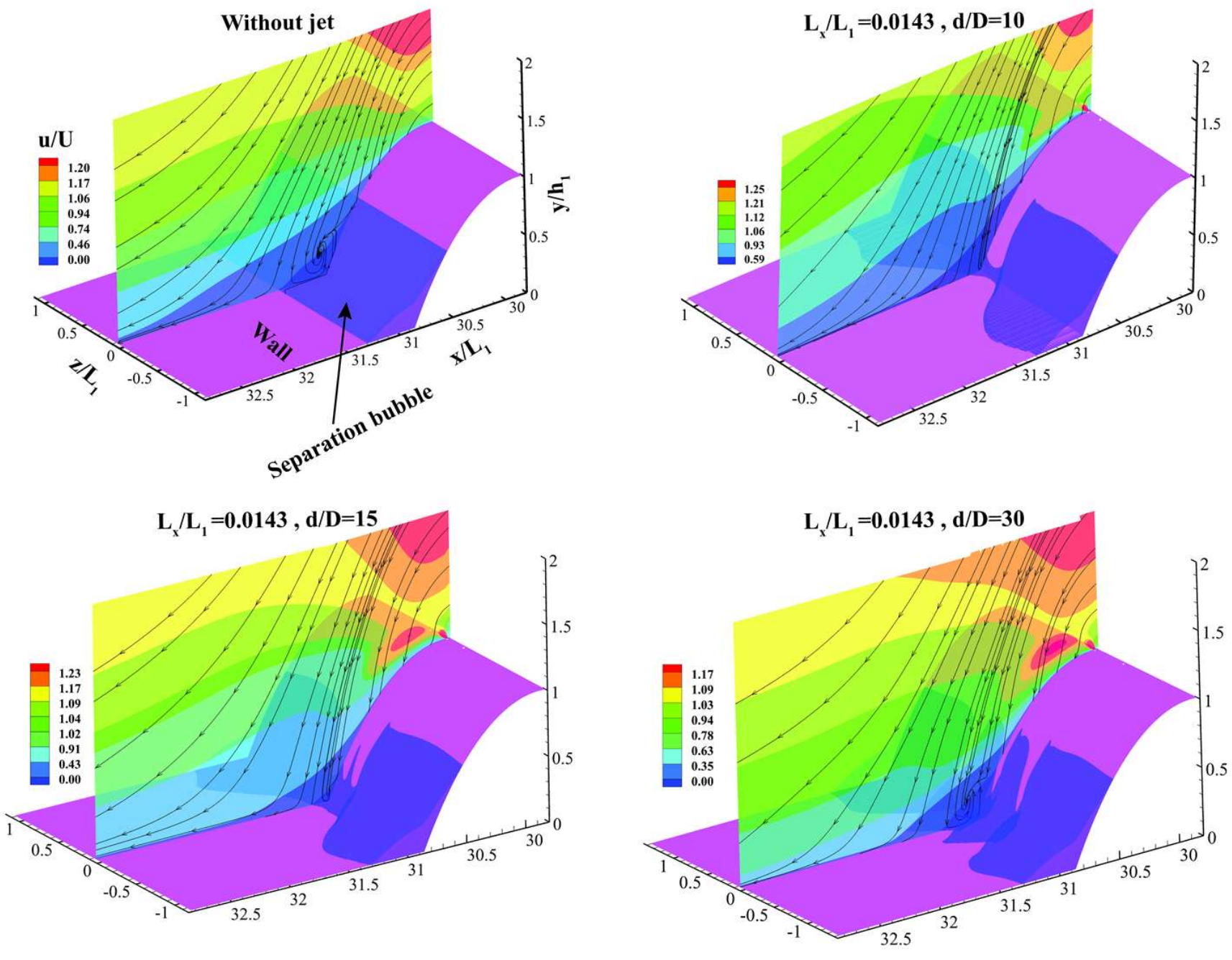

Figure 22: The separation bubble zone and velocity contours in the cross-plane as a function of the cross distance of jets from one another.

Velocity contours corresponding to different arrangements of jets are presented in Fig. 23. Again, velocity contours for the section in which $z/h_1$ = 0.0714 as well as the flow lines for the case without jet and the cases of positioning the jets in the arrangements with $L_x/L_1 = 0.0143$ and $d/D = 15$, $L_x/L_1 = 0.357$ and $d/D = 15$, and $L_x/L_1 = 1.43$ and $d/D = 15$ are illustrated. This figure shows that the jets' performance weakened when they were moved back. This fact is realized better by examining the separation zone when $L_x/L_1 = 1.43$. It is observed that separation has been prevented, but there are new zones in which the flow separation has risen enormously. In addition, the flow lines show that the flow rotation is removed in the case in which jets were employed.

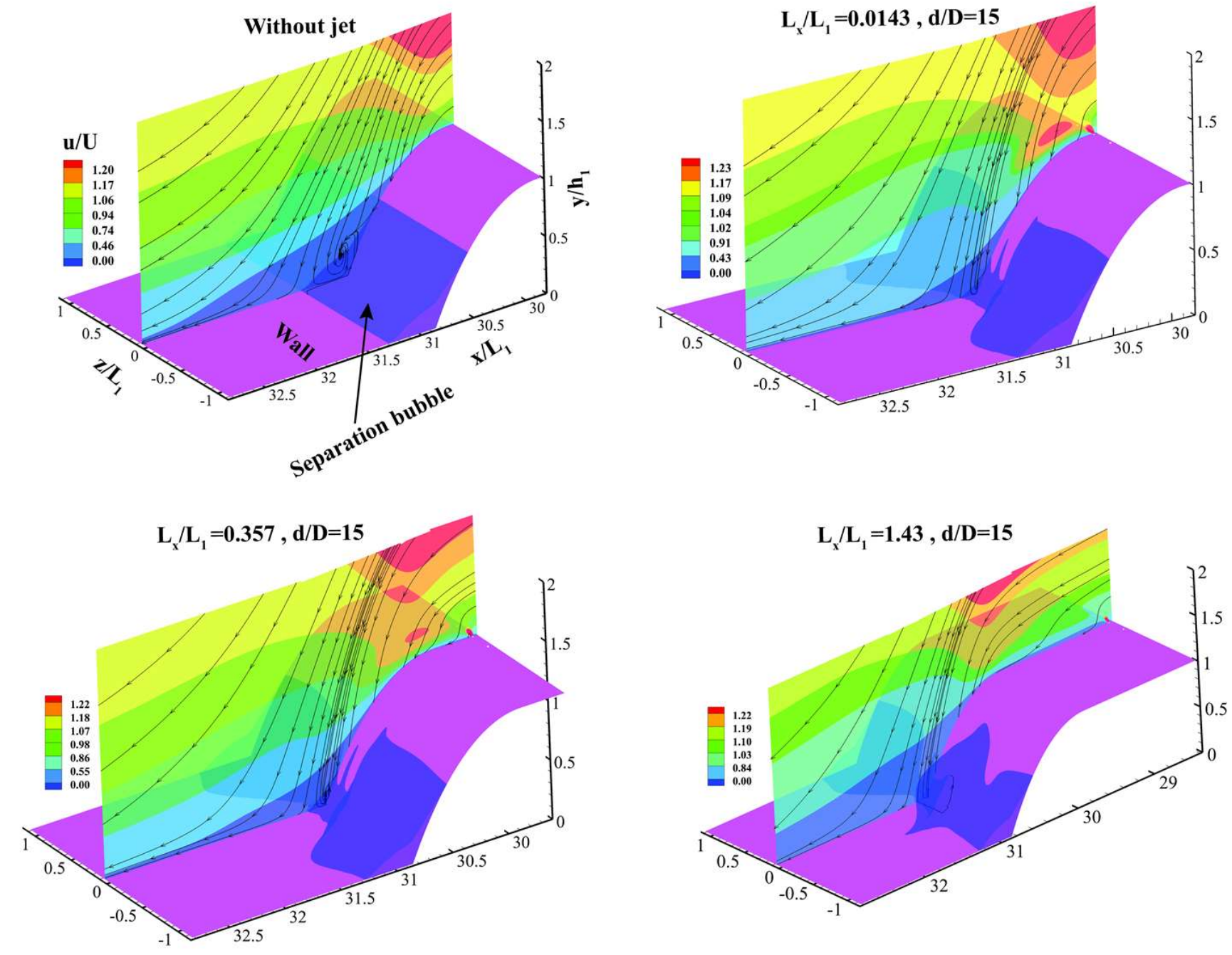

Figure 23: The separation bubble zone and velocity contours in the cross-plane as a function of the longitudinal coordinate of jets.

## *4. Conclusion*

This study analyzed the impacts of one and three jets on controlling a flow separation bubble over a ramp as an engineering problem. First, a single jet was placed over a ramp at different longitudinal locations to investigate the vanishing region of the separation bubble. The same study was then carried out with three jets, whose results were compared with those from the single jet. To this end, the results for the case of using no jet were compared to five cases with different positions of jets in an array in the form of (non-dimensional distance from top of the ramp, non-dimensional distance between two microjets) ($L_x/L_1 = 0.357$, $d/D = 15$), ($L_x/L_1 = 1.43$, $d/D = 15$), ($L_x/L_1 = 0.0143$, $d/D = 10$), ($L_x/L_1 = 0.0143$, $d/D = 15$), and ($L_x/L_1 = 0.0143$, $d/D = 30$). Moreover, velocity contours in longitudinal- and cross-planes, the flow lines, and the entrainment zone were investigated. The following results were obtained from the mentioned analyses.

- The results revealed that the flow separation started at about $x'$ = 0.75 (non-dimensional distance on the ramp from the top) and persisted up to about $x'$ = 1.44.
- Within its range of influence, the jets eliminated the separation bubble in this span. All three jets adequately prevented flow separation after the ramp. Nevertheless, they gave rise to even more separation in the upwash, generally the same for all three jets. There was a larger upwash and a stronger backflow for the further jet from the ramp. Therefore, it was concluded that using jets closer to the ramp was more effective in controlling the flow separation bubble.
- At $x'$ = 1, none of the jets could remove the separation bubble entirely.
- The jets prevented flow separation (dimensionless $\Delta z$ separation region) by 0.2 and 0.6 in the lateral direction at the beginning and end of the flow, respectively.
- The upwash regions of Jets 3 and 2 were larger than about 2 and 1.2 times of Jet 1, respectively.
- It is also observed that, as the vortices reached the ramp and the velocity along $y$ grew larger than that along $z$, these vortices dissipated at a distance less than 50 mm from the jet. Moreover, as their effects ran into the bed, two weak vortices were formed and moved along the flow. This phenomenon prevented vortex merging when a jet array was used.
- Using a jet array caused the separation bubble to grow laterally (along $z$) 2.1 and 2.75 times larger for $d/D$ = 10 and 15, respectively. Since there was no vortex merging, this region did not change in length or height compared to the case with a single jet.
- The jets prevented flow separation width by ($z/L_1$ = 0.2) and ($z/L_1$ = 0.6) in the lateral direction at the beginning and end of the separation flow region, respectively.
- The employment of jet managed to decrease the length of separation zone up to 78% and 71%, respectively, in the cases of ($L_x/L_1$ = 0.0143, $d/D$ = 10) and ($L_x/L_1$ = 0.0143, $d/D$ = 15).
- The upwash regions of Jets 3 and 2 were larger than and about 2 and 1.2 times, respectively, that of Jet 1.

## *Nomenclature*

| | |
|---|---|
| $t$ | Time |
| $U$ | Velocity |
| $g$ | Earth's gravitational acceleration |
| $P$ | Pressure |
| $k$ | Turbulence kinetic energy |
| $R$ | Eddy-viscosity parameter |
| $V_{jet}$ | Fluid microjet velocity |
| $d$ | Microjet diameter |
| $D$ | Distance between microjet |
| $L_x$ | Longitudinal distance of microjet from above the ramp |
| $L_1$ | Ramp length |
| $h_1$ | Ramp height |
| $\alpha$ | Jet pitch angle |
| $\beta$ | Jet skew angle with respect to the $x$-axis |
| $x^{'}$ | Longitudinal distance from above the ramp |
| $u$, $v$, and $w$ | $x$-, $y$-, and $z$-component of velocity |
| $u'$, $v'$, and $w'$ | $x$-, $y$-, and $z$-component of turbulent velocity |
| VR | Velocity ratio |
| $\rho$ | Density |
| $\mu$ | Dynamic viscosity |
| $\nu$ | Kinematic viscosity |
| $\nu_t$ | Turbulence viscosity |
| $\omega$ | $x$-component of vorticity |
| $\gamma$ | Intermittency |
| $Re_{\theta_c}$ | Critical Reynolds number |

## *Acknowledgments*